\batchmode
\makeatletter
\def\input@path{{"C:/Users/Windows/Desktop/Work Out/arxiv-Beltrami/"}}
\makeatother
\documentclass[english,aps,superscriptaddress,preprint]{revtex4-1}
\usepackage[T1]{fontenc}
\usepackage{verbatim}
\usepackage{amsmath}
\usepackage{graphicx}
\usepackage{babel}
\begin{document}
\preprint{USTC-ICTS/PCFT-22-02}
\title{Chandrasekhar-Kendall-Woltjer-Taylor state in a resistive plasma}
\author{Ze-Yu Zhai}
\email{zzymax@mail.ustc.edu.cn}

\affiliation{Interdisciplinary Center for Theoretical Study and Department of Modern
Physics, University of Science and Technology of China, Hefei, Anhui
230026, China}
\affiliation{Peng Huanwu Center for Fundamental Theory, Hefei, Anhui 230026, China}
\author{Yang-Guang Yang}
\email{mathmuse@ustc.edu.cn}

\affiliation{Interdisciplinary Center for Theoretical Study and Department of Modern
Physics, University of Science and Technology of China, Hefei, Anhui
230026, China}
\affiliation{Peng Huanwu Center for Fundamental Theory, Hefei, Anhui 230026, China}
\author{Xiao-Liang Xia}
\email{xiaxl@fudan.edu.cn}

\affiliation{Department of Physics and Center for Field Theory and Particle Physics,
Fudan University, Shanghai 200433, China}
\author{Qun Wang}
\email{qunwang@ustc.edu.cn}

\affiliation{Interdisciplinary Center for Theoretical Study and Department of Modern
Physics, University of Science and Technology of China, Hefei, Anhui
230026, China}
\affiliation{Peng Huanwu Center for Fundamental Theory, Hefei, Anhui 230026, China}
\begin{abstract}
We give a criterion for the Chandrasekhar-Kendall-Woltjer-Taylor (CKWT)
state in a resistive plasma. We find that the lowest momentum (longest
wavelength) of the initial helicity amplitudes of magnetic fields
are the key to the CKWT state which can be reached if one helicity
is favored over the other. This indicates that the imbalance between
two helicities at the lowest momentum or longest wavelength in the
initial conditions is essential to the CKWT state. A few examples
of initial conditions for helicity amplitudes are taken to support
the above statement both analytically and numerically.
\end{abstract}
\maketitle

\section{Introduction}

Many observations indicate that a magnetohydrodynamic (MHD) plasma
or a fluid can evolve into a special static state \citep{Robinson1969,Bodin1980,Ortolani1993,Sarff1997,Yagi1999,Sarff2003,Ding2004,Prager2005,Lorenzini2009},
in which a time-varying vector field is parallel to its curl, 
\begin{equation}
\boldsymbol{F}\times(\boldsymbol{\nabla}\times\boldsymbol{F})=0.
\end{equation}
This type of vector field is called the Beltrami field and was first
studied by Beltrami \citep{Beltrami1889}. In contrast, when the vector
field is orthogonal to its curl, 
\begin{equation}
\boldsymbol{F}\cdot(\boldsymbol{\nabla}\times\boldsymbol{F})=0,
\end{equation}
the field is called the complex lamellar field.

In the MHD plasma, the Beltrami field is just a force-free (magnetic)
field that was first discussed by Lust \citep{Lust1954} and Chandrasekhar
\citep{Chandrasekhar1957} in the context of cosmology. Then Chandrasekhar
and Woltjer \citep{Chandrasekhar1958,Woltjer1958} gave the first
analytical solution for such a force-free field. Such a state of the
MHD plasma is later called Chandrasekhar-Kendall-Woltjer-Taylor (CKWT)
state. The CKWT state satisfies the following equation for the magnetic
field, 
\begin{equation}
\boldsymbol{\nabla}\times\boldsymbol{B}=\lambda(\boldsymbol{r})\boldsymbol{B},
\end{equation}
where $\lambda(\boldsymbol{r})$ is a space-varying coefficient. If
$\lambda(\boldsymbol{r})=\lambda$ is a space constant, we call the
magnetic field is in a strong CKWT state. Otherwise, the magnetic
field is in a general CKWT state.

Some natural questions arise: can the CKWT state be reached? Under
what conditions can it be reached? Woltjer showed that the CKWT state
has the minimum magnetic energy at a fixed magnetic helicity \citep{Woltjer1958}.
As an invariant of plasma motion, magnetic helicity is associated
with the topological properties of the magnetic field lines and measures
them with the net twisting and braiding numbers \citep{Wells1969,Moffatt1978,Berger1984,Arnold1999}.
Later on, Taylor applied Woltjer's idea to a plasma with small electrical
resistance and found that Woltjer's condition is valid \citep{Taylor1974,Taylor1986}.
To provide a natural way to minimize the magnetic energy while keeping
the magnetic helicity fixed, Taylor speculated that the magnetic relaxation
is caused by small-scale turbulence \citep{Taylor1974,Taylor1986,Schnack2009}.
However, both experimental and theoretical studies did not give conclusive
evidence to support the hypothesis that the plasma relaxation should
be dominated by short-wavelength properties \citep{Robinson1969,Bodin1980,Caramana1983,Schnack1985,Strauss1985,Kusano1987,Ortolani1993,Holmes1988,Ho1991,Sarff1997,Yagi1999,Sarff2003,Diamond2003,Ding2004,Prager2005,Marrelli2005,Lorenzini2009}.
The idea that the fluctuations seem to have a global long-wavelength
structure is supported by extensive numerical simulations, which show
that the relaxation is caused by the long-wavelength instability and
nonlinear interaction.

To overcome the shortcoming of Taylor's theory, the relaxation theory
was developed using an infinite set of other approximate invariants
by different authors \citep{Bhattacharjee1980,Bhattacharjee1982}.
Another study on how to reach the CKWT state in the resistive plasmas
without Taylor's conjecture is proposed in Ref. \citep{Qin2012}.
Although the conditions in this work are not sufficient, the methods
are useful and have been applied in subsequent studies. The authors
of Ref. \citep{Hirono:2015rla} investigated the helicity evolution
of an expanding chiral plasma in magnetic fields with the chiral magnetic
effect \citep{Vilenkin:1980fu,Kharzeev:2007jp,Fukushima:2008xe} based
on an expansion of the fields in the vector spherical harmonics (VSH)
{[}for recent reviews of the chiral magnetic effect and related topics,
see, e.g. Refs. \citep{Kharzeev:2012ph,Kharzeev:2015znc}{]}. The
VSH method was later applied to study the CKWT state in a chiral plasma
with the chiral magnetic effect by some of us \citep{Xia:2016any},
and it is found that the chiral magnetic effect plays the role of
seed to the realization of the CKWT state.

A natural question arises: can the CKWT state be reached without the
chiral magnetic effect? In this paper, we are going to answer this
question by using the VSH method and a set of inequalities about magnetic
fields and vector potentials. We will propose a criterion for the
CKWT state, with which we find that the lowest momentum in the initial
helicity amplitudes of magnetic fields is the key to the CKWT state. 

The paper is organized as follows. In Section \ref{sec:General-discussions},
we will introduce the basic knowledge about the CKWT state. In Section
\ref{sec:observables-CKWT}, we will give the criterion for the CKWT
state through observables. In Section \ref{sec:Methods}, we will
introduce the VSH method to calculate the time evolution of these
observables. In Section \ref{sec:Approach-to-CKWT}, we will study
under which initial conditions the CKWT state can be reached. We will
summarize the main results of this paper in the final section.

\section{Basics of CKWT state \label{sec:General-discussions}}

We start from Maxwell equations, 
\begin{eqnarray}
\boldsymbol{\nabla}\times\boldsymbol{B} & = & \frac{\partial\boldsymbol{E}}{\partial t}+\boldsymbol{j},\label{eq:EM equs}\\
\boldsymbol{\nabla}\times\boldsymbol{E} & = & -\frac{\partial\boldsymbol{B}}{\partial t},\label{eq:EM equs-1}\\
\boldsymbol{\nabla}\cdot\boldsymbol{B} & = & 0,\label{eq:EM equs-2}\\
\boldsymbol{\nabla}\cdot\boldsymbol{E} & = & 0,\label{eq:EM equs-3}
\end{eqnarray}
where $\boldsymbol{E}$ and $\boldsymbol{B}$ are the electric and
magnetic field respectively. The current $\boldsymbol{j}$ reads 
\begin{equation}
\boldsymbol{j}=\sigma\boldsymbol{E},\label{eq:current eqns}
\end{equation}
where $\sigma$ is the electric conductivity. After taking a curl
of Eq. (\ref{eq:EM equs}), we obtain an evolution equation for the
magnetic field, 

\begin{eqnarray}
\frac{\partial^{2}}{\partial t^{2}}\boldsymbol{B}+\sigma\frac{\partial}{\partial t}\boldsymbol{B} & = & \nabla^{2}\boldsymbol{B}.\label{eq:EM evo 2nd}
\end{eqnarray}
In this paper we assume that $\sigma$ is a constant. We also assume
that terms of second-order time derivatives are much smaller than
those of first-order one, which is valid for a slowly time-varying
system. In this case, Eq. (\ref{eq:EM evo 2nd}) is reduced to 
\begin{equation}
\frac{\partial}{\partial t}\boldsymbol{B}=\eta\nabla^{2}\boldsymbol{B},\label{eq:EM evo 1st-sim}
\end{equation}
where $\eta=1/\sigma$ is the electrical resistance.

The authors of Ref. \citep{Qin2012} studied the general conditions
for the  CKWT state in a MHD plasma. It is helpful to introduce the
following inner products 
\begin{align}
W & =\left\langle \boldsymbol{B},\boldsymbol{B}\right\rangle =\int_{\Omega}\boldsymbol{B}^{2}d^{3}\boldsymbol{x},\nonumber \\
Q & =\left\langle \boldsymbol{A},\boldsymbol{A}\right\rangle =\int_{\Omega}\boldsymbol{A}^{2}d^{3}\boldsymbol{x},\nonumber \\
H & =\left\langle \boldsymbol{A},\boldsymbol{B}\right\rangle =\int_{\Omega}\boldsymbol{A}\cdot\boldsymbol{B}d^{3}\boldsymbol{x},\label{eq:wqh-bb-aa-ab}
\end{align}
where $W$ is the magnetic energy, $H$ is the magnetic helicity,
and $\Omega$ is the space volume. Using Eq. (\ref{eq:EM evo 1st-sim}),
we obtain 
\begin{align}
\frac{dQ}{dt} & =-2\eta\int_{\Omega}\boldsymbol{B}^{2}d^{3}\boldsymbol{x},\nonumber \\
\frac{dW}{dt} & =-2\eta\int_{\Omega}\boldsymbol{j}^{2}d^{3}\boldsymbol{x},\nonumber \\
\frac{dH}{dt} & =-2\eta\int_{\Omega}\boldsymbol{j}\cdot\boldsymbol{B}d^{3}\boldsymbol{x}.\label{eq:t-deriv-qwh}
\end{align}
After successively applying the Arithmetic Mean-Geometric Mean inequality
and Cauchy-Schwarz inequality, one can prove \citep{Qin2012}
\begin{eqnarray}
\frac{d}{dt}(WQ-H^{2}) & \leq & 4\eta\left[\int_{\Omega}\boldsymbol{A}\cdot\boldsymbol{B}d^{3}\boldsymbol{x}\int_{\Omega}\boldsymbol{j}\cdot\boldsymbol{B}d^{3}\boldsymbol{x}\right.\nonumber \\
 &  & \left.-\sqrt{\int_{\Omega}\boldsymbol{A}^{2}d^{3}\boldsymbol{x}\int_{\Omega}\boldsymbol{B}^{2}d^{3}\boldsymbol{x}\int_{\Omega}\boldsymbol{j}^{2}d^{3}\boldsymbol{x}\int_{\Omega}\boldsymbol{B}^{2}d^{3}\boldsymbol{x}}\right]\nonumber \\
 & \leq & 0.\label{eq:WQ-H2-dt}
\end{eqnarray}
The Cauchy-Schwartz inequality also gives the following inequality
\begin{equation}
WQ\geq\left(\int_{\Omega}|\boldsymbol{A}||\boldsymbol{B}|d^{3}\boldsymbol{x}\right)^{2}\geq\left(\int_{\Omega}\boldsymbol{A}\cdot\boldsymbol{B}d^{3}\boldsymbol{x}\right)^{2}=H^{2}.\label{eq:WQ-H2}
\end{equation}
Inequalities (\ref{eq:WQ-H2-dt}) and (\ref{eq:WQ-H2}) indicate that
the quantity $WQ-H^{2}$ is always positive and decreases with time
until the condition $\boldsymbol{B}=\lambda\boldsymbol{A}$ is reached,
in which $WQ-H^{2}$ is vanishing \citep{Qin2012}.

\section{Observables for CKWT State}

\label{sec:observables-CKWT}As shown in Eqs. (\ref{eq:WQ-H2-dt})
and (\ref{eq:WQ-H2}), $QW-H^{2}$ is always positive and decreases
with time unless $\boldsymbol{B}=\lambda\boldsymbol{A}$ is reached.
However, it is not sufficient to judge for the CKWT state only from
a decreasing $QW-H^{2}$, since it can decrease as the magnitudes
of $\boldsymbol{A}$ and $\boldsymbol{B}$ decrease while keeping
a fixed angle between them \citep{Chen2013}. The sufficient condition
for the CKWT state should be $\boldsymbol{B}$ and $\boldsymbol{A}$
are parallel. In this section, we propose to use the observable $WQ/H^{2}-1$
for the CKWT state provided $H\neq0$ and it is non-negative with
Cauchy-Schwartz inequality as we have shown in Section \ref{sec:General-discussions}.
We will show in this section that the condition for the CKWT state
should be 
\begin{align}
\frac{WQ}{H^{2}} & -1=\tan^{2}(\theta)\overset{t\rightarrow\infty}{\Longrightarrow}0,
\end{align}
where $\theta$ is an average angle between $\boldsymbol{A}$ and
$\boldsymbol{B}$ defined through $\left\langle \boldsymbol{A},\boldsymbol{B}\right\rangle ^{2}=\left\langle \boldsymbol{A},\boldsymbol{A}\right\rangle \left\langle \boldsymbol{B},\boldsymbol{B}\right\rangle \cos^{2}(\theta)$.

From $QW-H^{2}=QW\sin^{2}(\theta)$, we see that the sufficient condition
for the CKWT state is $\theta=0\text{ or }\pi$. It is more convenient
to introduce the quantity 
\begin{equation}
\tan^{2}\theta\equiv\frac{QW-H^{2}}{H^{2}}=\frac{\left\langle \boldsymbol{A},\boldsymbol{A}\right\rangle \left\langle \boldsymbol{B},\boldsymbol{B}\right\rangle -\left\langle \boldsymbol{A},\boldsymbol{B}\right\rangle ^{2}}{\left\langle \boldsymbol{A},\boldsymbol{B}\right\rangle ^{2}}.
\end{equation}
Assuming that $H\neq0$, the time rate of $QW-H^{2}$ can be expressed
as 
\begin{eqnarray}
\frac{d}{dt}(QW-H^{2}) & = & 2H^{2}\tan^{2}\theta(t)\left(\frac{d\ln|\tan\theta(t)|}{dt}+\frac{d\ln|H|}{dt}\right)\leq0,
\end{eqnarray}
with two contributions: the angular one and helicity one. We can prove
\begin{equation}
\frac{d\ln|\tan\theta(t)|}{dt}<0,\label{eq:angle-rate}
\end{equation}
for $t\rightarrow\infty$ in order to approach the CKWT state, which
means $\theta^{\prime}(t)<0$ for $\theta\in[0,\pi/2)$ and $\theta^{\prime}(t)>0$
for $\theta\in(\pi/2,\pi]$.

To prove that the necessary condition (\ref{eq:angle-rate}) is achievable,
we look at a simple case of the helicity time evolution. The long
time behaviors of $Q$ and $H$ lead to $d\ln|H|/dt\leq0$ when $t\rightarrow\infty$.
Then the condition (\ref{eq:angle-rate}) can be rewritten as 
\begin{equation}
\frac{1}{QW}\frac{d(QW)}{dt}<\frac{1}{H^{2}}\frac{dH^{2}}{dt}.\label{eq:angle-rate-1}
\end{equation}
We take a simple example of to illustrate the above condition. To
obtain an upper bound of the left-hand side of the above inequality,
we employ the Poincare inequality for the vector field $\boldsymbol{f}$
in following form \citep{Poincare1890}
\begin{equation}
\int_{\Omega}\boldsymbol{f}^{2}d^{3}\boldsymbol{x}\leq q_{\Omega}^{2}\int_{\Omega}(\boldsymbol{\nabla\times f})^{2}d^{3}\boldsymbol{x},
\end{equation}
where $q_{\Omega}$ is a Poincare constant associated with the space
volume $\Omega$. Then we obtain the upper bound as 
\begin{eqnarray}
\frac{1}{QW}\frac{d(QW)}{dt} & = & -2\eta\left(\frac{\int_{\Omega}\boldsymbol{B}^{2}d^{3}\boldsymbol{x}}{\int_{\Omega}\boldsymbol{A}^{2}d^{3}\boldsymbol{x}}+\frac{\int_{\Omega}\boldsymbol{j}^{2}d^{3}\boldsymbol{x}}{\int_{\Omega}\boldsymbol{B}^{2}d^{3}\boldsymbol{x}}\right)\nonumber \\
 & \leq & -2\eta(q_{\Omega}^{-2}+q_{\Omega}^{-2})=-4\eta q_{\Omega}^{-2}.
\end{eqnarray}
Since helicity is a topological quantity of plasma evolution, here
we postulate a tighter inequality than (\ref{eq:angle-rate-1})
\begin{equation}
-4\eta q_{\Omega}^{-2}<\frac{1}{H^{2}}\frac{dH^{2}}{dt},\label{eq:tight-cond}
\end{equation}
which can lead to (\ref{eq:angle-rate-1}) and (\ref{eq:angle-rate}).
So if the condition (\ref{eq:tight-cond}) is satisfied the angle
between $\boldsymbol{A}$ and \textbf{$\boldsymbol{B}$} decreases
with time. Furthermore, if $\theta(t)$ decreases fast enough, the
system will reach the CKWT state in a finite time. We still need to
know the time limit of $\tan^{2}\theta(t)$ in order to judge for
the CKWT state, which we will study in the next section.

\section{Methods}

\label{sec:Methods}To study the criteria for CKWT states, we need
to analyze the time evolution of $WQ/H^{2}$, it is convenient to
expand $W$, $Q$ and $H$ in (\ref{eq:wqh-bb-aa-ab}) as well as
their time rates in (\ref{eq:t-deriv-qwh}) on the VSH basis \citep{Jackson1999}.
The VSH basis functions are the eigenfunctions of the curl operator
in momentum space. They have been used to study the time evolution
of the magnetic helicity and the CKWT state in chiral plasma \citep{Hirono:2015rla,Xia:2016any}.

\subsection{VSH expansion}

We now expand $\boldsymbol{A}$ and $\boldsymbol{B}$ in terms of
the VSH basis functions $\boldsymbol{W}_{lm}^{s}(\boldsymbol{x},k)$
as 
\begin{eqnarray}
\boldsymbol{B}(t,\boldsymbol{x}) & = & \frac{1}{\pi}\sum_{l,m}\int_{0}^{\infty}dkk^{2}\left[\alpha_{lm}^{+}(t,k)\boldsymbol{W}_{lm}^{+}(\boldsymbol{x},k)+\alpha_{lm}^{-}(t,k)\boldsymbol{W}_{lm}^{-}(\boldsymbol{x},k)\right],\nonumber \\
\boldsymbol{A}(t,\boldsymbol{x}) & = & \frac{1}{\pi}\sum_{l,m}\int_{0}^{\infty}dkk\left[\alpha_{lm}^{+}(t,k)\boldsymbol{W}_{lm}^{+}(\boldsymbol{x},k)-\alpha_{lm}^{-}(t,k)\boldsymbol{W}_{lm}^{-}(\boldsymbol{x},k)\right],\label{eq:VSH-expansion}
\end{eqnarray}
where $\alpha_{lm}^{\pm}(t,k)$ denote the coefficients of the expansion,
and $\boldsymbol{W}_{lm}^{s}(\boldsymbol{x},k)$ (with $s=\pm$ being
the helicity) denote the complete set of eigenfunctions (vectors)
of the curl operator and are divergence-free 
\begin{eqnarray}
\boldsymbol{\nabla}\times\boldsymbol{W}_{lm}^{s}(\boldsymbol{x},k) & = & sk\boldsymbol{W}_{lm}^{s}(\boldsymbol{x},k),\nonumber \\
\boldsymbol{\nabla}\cdot\boldsymbol{W}_{lm}^{s}(\boldsymbol{x},k) & = & 0.\label{eq:curl-w}
\end{eqnarray}
In $\boldsymbol{W}_{lm}^{s}(\boldsymbol{x},k)$, $l=0,1,\cdots$ denotes
the orbital angular momentum quantum number, $m=-l,-l+1,\cdots,l$
denotes the magnetic quantum number, and $k\equiv|\boldsymbol{k}|$
is the norm of the momentum. The orthogonormality relations read 
\begin{eqnarray}
\int d^{3}\boldsymbol{x}\boldsymbol{W}_{l_{1}m_{1}}^{s_{1}}(\boldsymbol{x},k)\cdot\boldsymbol{W}_{l_{2}m_{2}}^{s_{2}}(\boldsymbol{x},k^{\prime}) & = & \frac{\pi}{k^{2}}\delta\left(k-k^{\prime}\right)\delta_{l_{1}l_{2}}\delta_{m_{1}m_{2}}\delta_{s_{1}s_{2}}.\label{eq:VSH orthorgonal}
\end{eqnarray}
To be specific, $\boldsymbol{W}_{lm}^{s}(\boldsymbol{x},k)$ can be
put into the form 
\begin{equation}
\boldsymbol{W}_{lm}^{s}(\boldsymbol{x},k)=\boldsymbol{T}_{lm}^{s}(\boldsymbol{x},k)+\frac{s}{k}\boldsymbol{\nabla}\times\boldsymbol{T}_{lm}^{s}(\boldsymbol{x},k),
\end{equation}
where $\boldsymbol{T}_{lm}^{s}(\boldsymbol{x},k)$ are toroidal fields
and can be expressed as a combination of spherical Bessel function
$j_{l}(kr)$ and spherical harmonic functions $Y_{lm}\left(\theta,\phi\right)$.

\subsection{Solving Maxwell equation}

Inserting Eq. (\ref{eq:VSH-expansion}) into Eq. (\ref{eq:EM evo 1st-sim}),
we obtain the evolution equation of the coefficients as 
\begin{eqnarray}
\frac{\partial}{\partial t}\alpha_{lm}^{\pm}(t,k) & = & -\eta k^{2}\alpha_{lm}^{\pm}(t,k),\label{eq:coe alpha equ}
\end{eqnarray}
where $\eta=1/\sigma$. Once $\alpha_{lm}^{\pm}(t,k)$ are obtained
by solving the above equation, the magnetic field $\boldsymbol{B}\left(t,\boldsymbol{x}\right)$
as a function of time is then known. The solutions of $\alpha_{lm}^{\pm}(t,k)$
are 
\begin{equation}
\alpha_{lm}^{\pm}(t,k)=e^{-\eta k^{2}t}\alpha_{lm}^{\pm}(0,k),
\end{equation}
where $\alpha_{lm}^{\pm}(0,k)$ are the values at the initial time
$t=0$. We need to calculate inner products of two fields as in Eq.
(\ref{eq:wqh-bb-aa-ab}). It is convenient to introduce positive-definite
functions $g_{\pm}(t,k)$ for the positive and negative helicity,
\begin{equation}
g_{\pm}(t,k)=\frac{1}{\pi}\sum_{lm}\left|\alpha_{lm}^{\pm}(t,k)\right|^{2}=e^{-2\eta k^{2}t}g_{\pm}(0,k),\label{eq:sol-g}
\end{equation}
where the initial values of $g_{\pm}(t,k)$ are $g_{\pm}(0,k)=(1/\pi)\sum_{lm}\left|\alpha_{lm}^{\pm}(0,k)\right|^{2}$.
In terms of $g_{\pm}(t,k)$, $W$, $Q$ and $H$ in (\ref{eq:wqh-bb-aa-ab})
can be put into the forms 
\begin{align}
W & =\int_{0}^{\infty}dkk^{2}\left[g_{+}(t,k)+g_{-}(t,k)\right],\nonumber \\
Q & =\int_{0}^{\infty}dk\left[g_{+}(t,k)+g_{-}(t,k)\right],\nonumber \\
H & =\int_{0}^{\infty}dkk\left[g_{+}(t,k)-g_{-}(t,k)\right],\label{eq:vsh-expand-wqh}
\end{align}
where we have used Eqs. (\ref{eq:VSH-expansion}-\ref{eq:VSH orthorgonal}).
We see in the above equations that $W$ and $Q$ are invariant or
symmetric under the interchange $g_{+}\leftrightarrow g_{-}$, while
$H$ is anti-symmetric under the the interchange $g_{+}\leftrightarrow g_{-}$.

\section{Approach to CKWT State}

\label{sec:Approach-to-CKWT}In this section, we will investigate
under what conditions the CKWT state is achieved.

\subsection{Special initial conditions}

We can explicitly express $WQ/H^{2}$ in terms of $g_{\pm}(0,k)$
using Eq. (\ref{eq:sol-g}), 
\begin{equation}
\frac{WQ}{H^{2}}=\frac{\int_{0}^{\infty}dk_{1}\int_{0}^{\infty}dk_{2}e^{-2\eta t(k_{1}^{2}+k_{2}^{2})}k_{1}^{2}\left[g_{+}(0,k_{1})+g_{-}(0,k_{1})\right]\left[g_{+}(0,k_{2})+g_{-}(0,k_{2})\right]}{\int_{0}^{\infty}dk_{1}\int_{0}^{\infty}dk_{2}e^{-2\eta t(k_{1}^{2}+k_{2}^{2})}k_{1}k_{2}\left[g_{+}(0,k_{1})-g_{-}(0,k_{1})\right]\left[g_{+}(0,k_{2})-g_{-}(0,k_{2})\right]},
\end{equation}
with its time derivative given by 
\begin{equation}
\frac{d}{dt}\left(\frac{WQ}{H^{2}}\right)=\frac{1}{H^{3}}\left(W^{\prime}QH+WQ^{\prime}H-2H^{\prime}WQ\right),
\end{equation}
where we have used the notation $X^{\prime}\equiv dX/dt$ with $X=W,Q,H$,
and the numerator and denominator have the explicit forms 
\begin{eqnarray}
 &  & W^{\prime}QH+WQ^{\prime}H-2H^{\prime}WQ\nonumber \\
 & = & \int_{0}^{\infty}dk_{1}\int_{0}^{\infty}dk_{2}\int_{0}^{\infty}dk_{3}e^{-2\eta t(k_{1}^{2}+k_{2}^{2}+k_{3}^{2})}k_{1}^{2}k_{3}\left(k_{1}^{2}+k_{2}^{2}-2k_{3}^{2}\right)\nonumber \\
 &  & \times\left[g_{+}(0,k_{1})+g_{-}(0,k_{1})\right]\left[g_{+}(0,k_{2})+g_{-}(0,k_{2})\right]\left[g_{+}(0,k_{3})-g_{-}(0,k_{3})\right],\nonumber \\
H^{3} & = & \int_{0}^{\infty}dk_{1}\int_{0}^{\infty}dk_{2}\int_{0}^{\infty}dk_{3}e^{-2\eta t(k_{1}^{2}+k_{2}^{2}+k_{3}^{2})}k_{1}k_{2}k_{3}\nonumber \\
 &  & \times\left[g_{+}(0,k_{1})-g_{-}(0,k_{1})\right]\left[g_{+}(0,k_{2})-g_{-}(0,k_{2})\right]\left[g_{+}(0,k_{3})-g_{-}(0,k_{3})\right].
\end{eqnarray}
In the following, we will look at the long-time behaviour of $WQ/H^{2}$
at $t\rightarrow\infty$ under some initial conditions. In the following
analysis and calculation, we use a typical length of the magnetic
field $L$ to scale the physical quantities, and we substitute $t\rightarrow t/L$,
$k\rightarrow kL$, $\eta\rightarrow\eta/L$ and $g_{\pm}\rightarrow g_{\pm}/L^{2}$,
so all re-scaled quantities are dimensionless.

\subsubsection{Delta functions}

First of all, we consider an ideal case of initial functions $g_{\pm}\left(0,k\right)$
with two different discrete momentum values $a$ and $b$ 
\begin{eqnarray}
g_{+}(0,k) & = & a\delta(k-a)+\frac{b}{2}\delta(k-b),\nonumber \\
g_{-}(0,k) & = & \frac{b}{2}\delta(k-b).
\end{eqnarray}
We see that the positive helicity part has two momentum values while
the negative helicity part has only one value. It is easy to obtain
\begin{equation}
\frac{WQ}{H^{2}}-1=\frac{1}{a^{4}}\left[\left(a^{3}b+ab^{3}\right)e^{2\left(a^{2}-b^{2}\right)\eta t}+b^{4}e^{4\left(a^{2}-b^{2}\right)\eta t}\right],
\end{equation}
with the $t\rightarrow\infty$ limit 
\begin{equation}
\lim_{t\rightarrow\infty}\frac{WQ}{H^{2}}-1\rightarrow\begin{cases}
0 & a<b\\
\infty & a>b
\end{cases}.
\end{equation}
We can see that only when $a<b$ the CKWT state can be achieved at
$t\rightarrow\infty$. In this case, helicity is dominated by the
low momentum mode. On the other hand, if the high momentum mode is
dominant, the CKWT state cannot be reached. Nevertheless, since the
delta function is not mathematically well-defined and should be replaced
by more physical initial conditions, this simple case still provides
a clue to more general conditions.

\begin{comment}
The derivation
\begin{eqnarray*}
\frac{WQ}{H^{2}}-1 & = & \frac{1}{k_{1}^{2}}\left(2e^{2(k_{1}^{2}-k_{2}^{2})t}+1\right)\left(k_{1}^{2}+2e^{2(k_{1}^{2}-k_{2}^{2})t}k_{2}^{2}\right)-1\\
 & = & \frac{1}{k_{1}^{2}}\left[2(k_{1}^{2}+k_{2}^{2})e^{2(k_{1}^{2}-k_{2}^{2})t}+4k_{2}^{2}e^{4(k_{1}^{2}-k_{2}^{2})t}+k_{1}^{2}\right]-1\\
 & = & \frac{2}{k_{1}^{2}}\left[2k_{2}^{2}e^{4(k_{1}^{2}-k_{2}^{2})t}+(k_{1}^{2}+k_{2}^{2})e^{2(k_{1}^{2}-k_{2}^{2})t}\right]
\end{eqnarray*}
\end{comment}

\subsubsection{Two-band functions}

\begin{figure}
\caption{\label{fig:WQ-H2-two-band}Numerical results for $\tan^{2}(\theta)$
with two-band initial conditions for different set of amplitude values.
The parameters are: $\eta=0.1$, $k_{d1}=0\;\mathrm{or}\;1$, $k_{d2}=2$,
$k_{d3}=4$. A: $d_{1}^{+}=1$, $d_{2}^{+}=1/4$, $d_{1}^{-}=0$,
$d_{2}^{-}=1/4$, B: $d_{1}^{+}=1/4$, $d_{2}^{+}=1$, $d_{1}^{-}=0$,
$d_{2}^{-}=1/4$, C: $d_{1}^{+}=1/4$, $d_{2}^{+}=1$, $d_{1}^{-}=1/4$,
$d_{2}^{-}=0$, D: $d_{1}^{+}=1$, $d_{2}^{+}=1/4$, $d_{1}^{-}=1/4$,
$d_{2}^{-}=0$, E: $d_{1}^{+}=1$, $d_{2}^{+}=1/4$, $d_{1}^{-}=1/3$,
$d_{2}^{-}=1/3$. (a) $k_{d1}=1$. In case A and B, $\tan^{2}(\theta)$
tends to zero as $t\rightarrow\infty$ indicating that the CKWT state
can be reached, while $\tan^{2}(\theta)$ tends to infinity, $16/9$
and $3$ as $t\rightarrow\infty$ in case C, D and E, respectively,
which indicates that the CKWT state cannot be reached. (b) $k_{d1}=0$.
The CKWT state cannot be reached in all cases.}

\centering{}\includegraphics[scale=0.22]{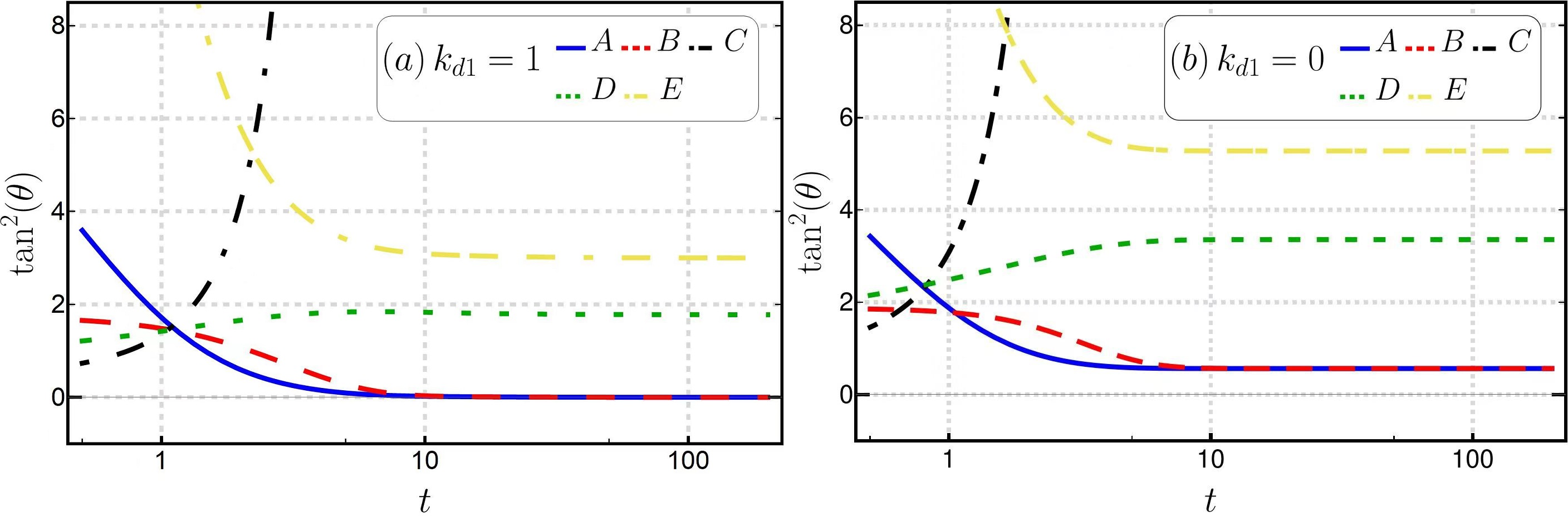}
\end{figure}

As a more general case than delta-functions, we consider Heaviside
step functions for $g_{\pm}(0,k)$ with two bands (the lower momentum
band and higher momentum band), 
\begin{align}
g_{+}(0,k) & =\begin{cases}
d_{1}^{+}, & \mathrm{for}\;\;k_{d1}\leq k<k_{d2}\\
d_{2}^{+}, & \mathrm{for}\;\;k_{d2}\leq k<k_{d3}
\end{cases},\\
g_{-}(0,k) & =\begin{cases}
d_{1}^{-}, & \mathrm{for}\;\;k_{d1}\leq k<k_{d2}\\
d_{2}^{-}, & \mathrm{for}\;\;k_{d2}\leq k<k_{d3}
\end{cases},
\end{align}
where $k\geq0$, $k_{d3}>k_{d2}>k_{d1}\geq0$, $d_{1}^{+}+d_{1}^{-}>0$
and $d_{i}^{\pm}\geq0$ for $i=1,2$. We can verify the following
limit when $t$ is sent to infinity, 
\begin{equation}
\lim_{t\rightarrow\infty}\frac{WQ}{H^{2}}-1\rightarrow\begin{cases}
\frac{\left(d_{1}^{+}+d_{1}^{-}\right)^{2}}{\left(d_{1}^{+}-d_{1}^{-}\right)^{2}}-1, & \mathrm{for}\;\;k_{d1}>0\\
\frac{\pi}{2}\frac{\left(d_{1}^{+}+d_{1}^{-}\right)^{2}}{\left(d_{1}^{+}-d_{1}^{-}\right)^{2}}-1, & \mathrm{for}\;\;k_{d1}=0
\end{cases}.
\end{equation}
We see that such a limit is determined by the amplitudes of the lower
momentum bands. The conditions for the CKWT state would be 
\begin{equation}
\Delta\equiv\frac{\left(d_{1}^{+}+d_{1}^{-}\right)^{2}}{\left(d_{1}^{+}-d_{1}^{-}\right)^{2}}=\left\{ \begin{array}{ll}
1, & \mathrm{for}\;\;k_{d1}>0\\
2/\pi, & \mathrm{for}\;\;k_{d1}=0
\end{array}\right..
\end{equation}
Because $d_{1}^{\pm}\geq0$ and $d_{1}^{+}+d_{1}^{-}>0$, $\Delta$
must not be less than 1 or $\Delta\geq1$, so $\Delta$ cannot be
$2/\pi$ for the case $k_{d1}=0$, which means that the CKWT state
cannot be reached for $k_{d1}=0$. For $k_{d1}>0$, the CKWT state
can be reached if and only if either of $d_{1}^{+}$ or $d_{1}^{-}$
is vanishing. This observation can be verified by the numerical results
in Fig. \ref{fig:WQ-H2-two-band} for different sets of values of
$d_{1}^{\pm}$ and $d_{2}^{\pm}$.

\subsubsection{Multi-band functions}

\begin{figure}[h]
\caption{\label{fig:WQ-H-multi-band}Numerical results for $\tan^{2}(\theta)$
with multi-band initial conditions for different sets of amplitude
values. The parameters are: $\eta=0.1$, $k_{d1}=0\;\mathrm{or}\;1$,
$k_{d2}=2$, $k_{d3}=5$, $k_{d4}=10$, $k_{d5}=16$. Case A: $\{d_{1}^{+},\text{\ensuremath{d_{2}^{+}}},d_{3}^{+},d_{4}^{+}\}=\{1,1,1,0\}$,
$\{d_{1}^{-},\text{\ensuremath{d_{2}^{-}}},d_{3}^{-},d_{4}^{-}\}=\{0,1/4,0,1\}$;
Case B: $\{d_{1}^{+},\text{\ensuremath{d_{2}^{+}}},d_{3}^{+},d_{4}^{+}\}=\{1/4,2,1,0\}$,
$\{d_{1}^{-},\text{\ensuremath{d_{2}^{-}}},d_{3}^{-},d_{4}^{-}\}=\{0,3,0,1/4\}$;
Case C: $\{d_{1}^{+},\text{\ensuremath{d_{2}^{+}}},d_{3}^{+},d_{4}^{+}\}=\{1/4,1,0,1\}$,
$\{d_{1}^{-},\text{\ensuremath{d_{2}^{-}}},d_{3}^{-},d_{4}^{-}\}=\{1/4,4,0,1\}$;
Case D: $\{d_{1}^{+},\text{\ensuremath{d_{2}^{+}}},d_{3}^{+},d_{4}^{+}\}=\{1,2,1,0\}$,
$\{d_{1}^{-},\text{\ensuremath{d_{2}^{-}}},d_{3}^{-},d_{4}^{-}\}=\{1/4,3,1,1\}$;
Case E: $\{d_{1}^{+},\text{\ensuremath{d_{2}^{+}}},d_{3}^{+},d_{4}^{+}\}=\{1,1/4,1,0\}$,
$\{d_{1}^{-},\text{\ensuremath{d_{2}^{-}}},d_{3}^{-},d_{4}^{-}\}=\{1/3,1/4,0,1\}$.
(a) $k_{d1}=1$. In case A and B, the CKWT state can be reached as
$t\rightarrow\infty$. In other cases the CKWT state cannot be reached.
(b) $k_{d1}=0$. The CKWT state cannot be reached for all cases.}

\centering{}\includegraphics[scale=0.22]{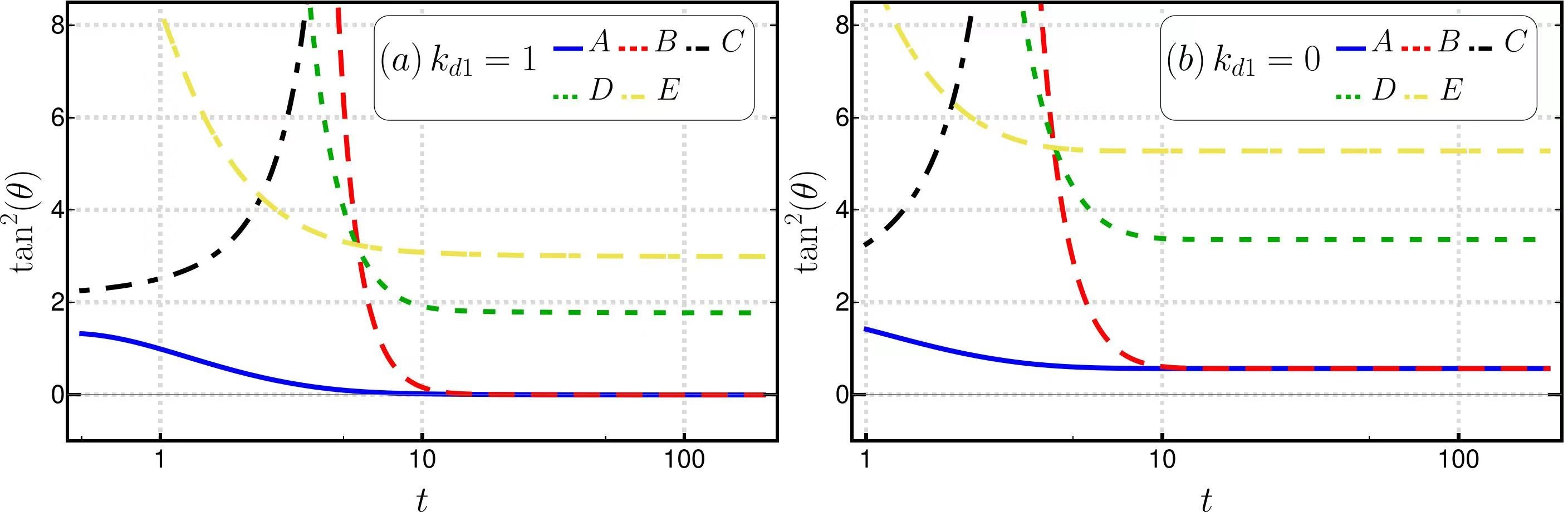}
\end{figure}

We now generalize two-steps functions to multi-steps functions, 
\begin{align}
g_{+}\left(0,k\right) & =\begin{cases}
d_{1}^{+}, & \mathrm{for}\;\;k_{d1}\leq k<k_{d2}\\
d_{2}^{+}, & \mathrm{for}\;\;k_{d2}\leq k<k_{d3}\\
... & ...\\
d_{n}^{+}, & \mathrm{for}\;\;k_{d(n)}\leq k<k_{d(n+1)}
\end{cases},
\end{align}

\begin{equation}
g_{-}\left(0,k\right)=\begin{cases}
d_{1}^{-}, & \mathrm{for}\;\;k_{d1}\leq k<k_{d2}\\
d_{2}^{-}, & \mathrm{for}\;\;k_{d2}\leq k<k_{d3}\\
... & ...\\
d_{n}^{-}, & \mathrm{for}\;\;k_{d(n)}\leq k<k_{d(n+1)}
\end{cases},
\end{equation}
where $k\geq0$, $k_{d(n+1)}>k_{d(n)}>...>k_{d2}>k_{d1}\geq0$, $d_{1}^{+}+d_{1}^{-}>0$
and $d_{i}^{\pm}\geq0$ for $i=1,2,...,n$. The result is similar
to the case of two-bands functions, 
\begin{equation}
\lim_{t\rightarrow\infty}\frac{WQ}{H^{2}}-1\rightarrow\begin{cases}
\frac{\left(d_{1}^{+}+d_{1}^{-}\right)^{2}}{\left(d_{1}^{+}-d_{1}^{-}\right)^{2}}-1, & \mathrm{for}\;\;k_{d1}>0\\
\frac{\pi}{2}\frac{\left(d_{1}^{+}+d_{1}^{-}\right)^{2}}{\left(d_{1}^{+}-d_{1}^{-}\right)^{2}}-1, & \mathrm{for}\;\;k_{d1}=0
\end{cases}.\label{eq:limit-multi-steps}
\end{equation}
We see that the limit for $WQ/H^{2}$ is also determined by the amplitudes
of the lowest bands. Similar to the analysis in the previous subsection
that the CKWT state can only be reached for $k_{d1}>0$ under the
condition that either of $d_{1}^{+}$ or $d_{1}^{-}$ is vanishing.
This observation can be verified by the numerical results in Fig.
\ref{fig:WQ-H-multi-band} for different sets of values of $d_{i}^{\pm}$
for $i=1,\cdots,4$.

\subsection{Analytic functions as initial conditions}

Based on the previous discussion about step functions as initial conditions,
it is natural to generalize it to the limit of infinitely small intervals,
i.e. analytic functions. As discussed in the previous subsection,
the starting point of the integral can make a difference in the limit
of $WQ/H^{2}$. In this subsection, we will do the same thing by distinguishing
two cases: $a>0$ and $a=0$ for the starting point of the integral
range $[a,\infty)$.

\subsubsection{Integration range $[a,\infty)$ with $a>0$}

The physical quantities in our consideration are all in the integrated
form 
\begin{equation}
X=\int_{a}^{\infty}dkk^{n}e^{-2\eta tk^{2}}f(k),
\end{equation}
where the lower bound $a$ of the integration range is a positive
number, and $f(k)$ is an analytic function, meaning that the Taylor
expansion is valid at any value $k_{0}$ in the range $[a,\infty)$,
\begin{equation}
f(k)=\sum_{n=0}^{\infty}\frac{1}{n!}\left(k-k_{0}\right)^{n}f^{(n)}(k_{0}).
\end{equation}
For $W$, $Q$ and $H$ which we are considering in this paper, $f(k)$
can be either $\phi(k)$ or $\varphi(k)$, 
\begin{eqnarray}
\phi(k) & = & g_{+}(0,k)+g_{-}(0,k),\nonumber \\
\varphi(k) & = & g_{+}(0,k)-g_{-}(0,k).\label{eq:phi-beta}
\end{eqnarray}
So $W$, $Q$ and $H$ can be put into the forms, 
\begin{eqnarray}
W & = & \int_{a}^{\infty}dkk^{2}e^{-2\eta tk^{2}}\left[\sum_{n=0}^{\infty}\frac{1}{n!}\phi^{(n)}(a)\left(k-a\right)^{n}\right],\nonumber \\
Q & = & \int_{a}^{\infty}dke^{-2\eta tk^{2}}\left[\sum_{n=0}^{\infty}\frac{1}{n!}\phi^{(n)}(a)\left(k-a\right)^{n}\right],\nonumber \\
H & = & \int_{a}^{\infty}dkke^{-2\eta tk^{2}}\left[\sum_{n=0}^{\infty}\frac{1}{n!}\varphi^{(n)}\left(a\right)\left(k-a\right)^{n}\right].
\end{eqnarray}
Then we obtain the long time limit 
\begin{equation}
\lim_{t\rightarrow\infty}\frac{WQ}{H^{2}}-1\rightarrow\left[\frac{g_{+}^{(i)}(0,a)+g_{-}^{(i)}(0,a)}{g_{+}^{(i)}(0,a)-g_{-}^{(i)}(0,a)}\right]^{2}-1,\label{eq:wq-h2-continuous}
\end{equation}
where one can explicitly define the derivative index $i$ ($i\geq0$)
with the following two cases: (a) $i=\min(i_{+},i_{-})$ if $g_{+}(0,k)$
and $g_{-}(0,k)$ are all non-vanishing, where $i_{s}$ is the index
denote for a lowest order $i_{s}$-th derivative that makes $g_{s}^{(i_{s})}(0,a)$
non-vanishing with $s=\pm$, respectively; (b) If one of $g_{+}(0,k)$
and $g_{-}(0,k)$ is vanishing, for example, $g_{+}(0,k)=0$, then
for a lowest order $i$-th derivative $g_{-}^{(i)}(0,a)\neq0$. So
is the case $g_{-}(0,k)=0$.

We see that the CKWT can be reached at $t\rightarrow\infty$ if and
only if either $g_{+}^{(i)}(0,a)$ or $g_{-}^{(i)}(0,a)$ is vanishing.
The proof of the result (\ref{eq:wq-h2-continuous}) is given in Appendix
\ref{sec:proof}.

\subsubsection{Integration range $[0,\infty)$}

In this section we consider the integration range $[0,\infty)$, in
which $g_{+}(0,k)$ and $g_{-}(0,k)$ are analytic functions and can
be expanded in a Taylor expansion. In the following we use the shorthand
notation $g_{\pm}(k)$ for $g_{\pm}(0,k)$.

At $k=0$, the Taylor expansion of $\phi(k)$ and $\varphi(k)$ in
Eq. (\ref{eq:phi-beta}) reads 
\begin{eqnarray}
\phi(k) & = & \phi(0)+\phi^{\prime}(0)k+\frac{1}{2}\phi^{\prime\prime}(0)k^{2}+...,\nonumber \\
\varphi(k) & = & \varphi(0)+\varphi^{\prime}(0)k+\frac{1}{2}\varphi^{\prime\prime}(0)k^{2}+....
\end{eqnarray}
By switching the order of the summation and the integration, one can
calculate the integration of every term, 
\begin{align}
\frac{1}{n!}\int_{0}^{\infty}e^{-2\eta tk^{2}}\phi^{(n)}(0)k^{p+n}dk & \propto t^{-(1+p+n)/2},\nonumber \\
\frac{1}{n!}\int_{0}^{\infty}e^{-2\eta tk^{2}}\varphi^{(n)}(0)k^{1+n}dk & \propto t^{-(2+n)/2},
\end{align}
where $p=0,2$ for $Q,W$ respectively. One can get rid of the term
$t^{-(1+p+n)/2}$ for $n>i$ when $t$ goes to infinity, where $i$
is the the derivative index denote for a lowest order $i$-th ($i\geq0$)
derivative that makes $\phi^{(i)}(0)$ non-vanishing.

Similarly, we obtains the final result 
\begin{equation}
\lim_{t\rightarrow\infty}\frac{WQ}{H^{2}}-1\rightarrow\left[\frac{g_{+}^{(i)}(0)+g_{-}^{(i)}(0)}{g_{+}^{(i)}(0)-g_{-}^{(i)}(0)}\right]^{2}\frac{\Gamma\left(\frac{3+i}{2}\right)\Gamma\left(\frac{1+i}{2}\right)}{\Gamma\left(\frac{2+i}{2}\right)\Gamma\left(\frac{2+i}{2}\right)}-1.\label{eq:zero-infinity-range}
\end{equation}
One can prove that the first factor in the right-hand-side of Eq.
(\ref{eq:zero-infinity-range}) is always larger than or equal to
1, and it is 1 if and only if either $g_{+}^{(i)}(0)=0$ or $g_{-}^{(i)}(0)=0$.
The second factor can be easily proved to be larger than 1. As a consequence,
$WQ/H^{2}$ is always larger than 1 and will not reach $1$ as $t\rightarrow\infty$,
so the CKWT state cannot be reached in this case.

\begin{figure}[h]
\caption{\label{fig:continuous-func}Numerical results for $\tan^{2}(\theta)$
for different continuous functions as initial conditions. The parameters
are: $\eta=1$, $g_{1+}(0,k)=\cos(k-d)+1$, $g_{2+}(0,k)=e^{-(k-d)}$,
$g_{3+}(0,k)=f(k)e^{-k}$, $g_{1-}(0,k)=(k-d)e^{-(k-d)}$, $g_{2-}(0,k)=|\sin(k-d)|$,
$g_{3-}(0,k)=(k^{3}/10)e^{-k^{2}}$. (a) Case A: $g_{\pm}(0,k)=g_{1\pm}(0,k)$,
$d=0.1$; Case B: $g_{\pm}(0,k)=g_{2\pm}(0,k)$, $d=0.1$; Case C:
$g_{\pm}(0,k)=g_{1\pm}(0,k)$, $d=0$; Case D: $g_{\pm}(0,k)=g_{2\pm}(0,k)$,
$d=0$. The integral range of them are all $[d,\infty)$. In case
A and B, the CKWT state can be reached as $t\rightarrow\infty$. In
other cases the CKWT state cannot be reached. (b) Case E: $g_{\pm}(0,k)=g_{3\pm}(0,k)$,
$f(k)=1$; Case F: $g_{\pm}(0,k)=g_{3\pm}(0,k)$, $f(k)=k$; Case
G: $g_{\pm}(0,k)=g_{3\pm}(0,k)$, $f(k)=k^{2}/2$. The integral range
of them are all $[0,\infty)$. The CKWT state cannot be reached for
all cases of (b). The black dashed lines in figures show the analytical
results for each curve with $l1(0)=0$, $l2(0)=\Gamma(3/2)\Gamma(1/2)/(\Gamma(1)\Gamma(1))-1\approx0.571$,
$l2(1)=\Gamma(2)\Gamma(1)/(\Gamma(3/2)\Gamma(3/2))-1\approx0.273$
and $l2(2)=\Gamma(5/2)\Gamma(3/2)/(\Gamma(2)\Gamma(2))-1\approx0.178$.}

\centering{}\includegraphics[scale=0.38]{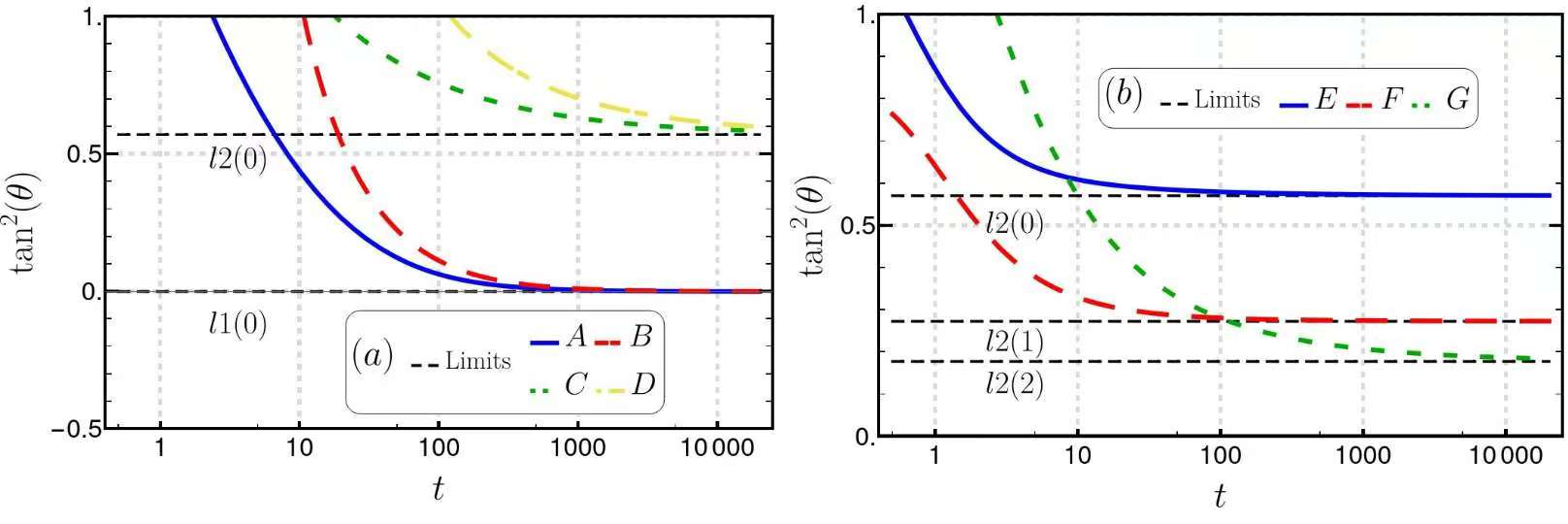}
\end{figure}

We see that the result for $a>0$ cannot be simply extended to that
for $a=0$ by taking the limit $a\rightarrow0$. The analytical result
can be verified numerically as presented in Fig. \ref{fig:continuous-func}.

\subsection{Special non-analytic function}

For non-analytic functions as initial conditions, it is difficult
to reach a similar conclusion as in previous sections. We can only
take an example and carry out our numerical calculations. We consider
the following function as the initial condition 
\begin{equation}
g_{+}(0,k)=\begin{cases}
e^{-1/k^{2}}, & k>0\\
0, & k=0
\end{cases}.
\end{equation}
What is special for this function is that it has infinite order of
derivatives at $k=0$ which are vanishing. Thus $g_{+}(0,k)$ is non-analytic
at $k=0$ and cannot be expanded into a Taylor series because zero
is the essential singularity in the complex domain. For convenience,
we assume $g_{-}(0,k)=0$ and the integration range is $[0,\infty)$,
then we can calculate $WQ/H^{2}$ directly and find the long time
limit with the second kind modified Bessel function $K_{\nu}(z)$ 

\begin{eqnarray}
\lim_{t\rightarrow\infty}\frac{WQ}{H^{2}}-1=\lim_{t\rightarrow\infty}\frac{e^{-4\sqrt{2\eta t}}\pi\left(1+2\sqrt{2\eta t}\right)}{16\eta t\left[K_{1}(2\sqrt{2\eta t})\right]^{2}}-1 & \rightarrow & 0.
\end{eqnarray}
We see that under this condition the CKWT state can be reached.

\section{Conclusion}

We have studied how the Chandrasekhar-Kendall-Woltjer-Taylor (CKWT)
state can be reached in the time evolution of a resistive plasma.
We propose a criterion for the CKWT state as the destination of the
time evolution, $\lim_{t\rightarrow\infty}\tan^{2}(\theta)\rightarrow0$,
where $\theta$ is the average angle between the magnetic field and
the vector potential. We find that the initial conditions for the
helicity amplitudes of the magnetic field and the vector potential
are essential to the CKWT state. Our analysis is based on an expansion
in the vector spherical harmonics for magnetic fields and vector potentials.

The asymptotic form of $\tan^{2}(\theta)$ is dominated by the lowest
momentum $k_{\mathrm{min}}$ of the initial helicity amplitudes $g_{\pm}(0,k)$
as functions of the scalar momentum $k$. For those initial helicity
amplitudes that can be expanded into a Taylor series, the CKWT state
cannot be reached if $k_{\mathrm{min}}=0$, while it can be reached
for $k_{\mathrm{min}}>0$ if and only if either $g_{+}^{(i)}(0,k_{\mathrm{min}})=0$
or $g_{-}^{(i)}(0,k_{\mathrm{min}})=0$ with the lowest $i$-th non-zero
derivative ($i\geq0$) explained in Section \ref{sec:Approach-to-CKWT}.
In other words, the CKWT state can be reached if one helicity is favored
over the other at the lowest momentum in the initial helicity amplitudes
of the magnetic field. This indicates that the imbalance between two
helicities at the lowest momentum (longest wavelength) in the initial
helicity amplitudes is the key factor for the CKWT state.

%done. 2022.01.23

\textit{Acknowledgments}. Z.Y.Z., Y.G.Y. and Q.W. are supported in
part by National Natural Science Foundation of China (NSFC) under
Grant Nos. 12135011, 11890713 (a subgrant of 11890710) and 12047502,
and by the Strategic Priority Research Program of Chinese Academy
of Sciences under Grant No. XDB34030102.

\bibliographystyle{apsrev4-1}
\bibliography{3C__Users_Windows_Desktop_Work_Out_arxiv-Beltrami_Beltrami-manuscript}

\begin{thebibliography}{42}%
\makeatletter
\providecommand \@ifxundefined [1]{%
 \@ifx{#1\undefined}
}%
\providecommand \@ifnum [1]{%
 \ifnum #1\expandafter \@firstoftwo
 \else \expandafter \@secondoftwo
 \fi
}%
\providecommand \@ifx [1]{%
 \ifx #1\expandafter \@firstoftwo
 \else \expandafter \@secondoftwo
 \fi
}%
\providecommand \natexlab [1]{#1}%
\providecommand \enquote  [1]{``#1''}%
\providecommand \bibnamefont  [1]{#1}%
\providecommand \bibfnamefont [1]{#1}%
\providecommand \citenamefont [1]{#1}%
\providecommand \href@noop [0]{\@secondoftwo}%
\providecommand \href [0]{\begingroup \@sanitize@url \@href}%
\providecommand \@href[1]{\@@startlink{#1}\@@href}%
\providecommand \@@href[1]{\endgroup#1\@@endlink}%
\providecommand \@sanitize@url [0]{\catcode `\\12\catcode `\$12\catcode
  `\&12\catcode `\#12\catcode `\^12\catcode `\_12\catcode `\%12\relax}%
\providecommand \@@startlink[1]{}%
\providecommand \@@endlink[0]{}%
\providecommand \url  [0]{\begingroup\@sanitize@url \@url }%
\providecommand \@url [1]{\endgroup\@href {#1}{\urlprefix }}%
\providecommand \urlprefix  [0]{URL }%
\providecommand \Eprint [0]{\href }%
\providecommand \doibase [0]{http://dx.doi.org/}%
\providecommand \selectlanguage [0]{\@gobble}%
\providecommand \bibinfo  [0]{\@secondoftwo}%
\providecommand \bibfield  [0]{\@secondoftwo}%
\providecommand \translation [1]{[#1]}%
\providecommand \BibitemOpen [0]{}%
\providecommand \bibitemStop [0]{}%
\providecommand \bibitemNoStop [0]{.\EOS\space}%
\providecommand \EOS [0]{\spacefactor3000\relax}%
\providecommand \BibitemShut  [1]{\csname bibitem#1\endcsname}%
\let\auto@bib@innerbib\@empty
%</preamble>
\bibitem [{\citenamefont {Robinson}(1969)}]{Robinson1969}%
  \BibitemOpen
  \bibfield  {author} {\bibinfo {author} {\bibfnamefont {D.~C.}\ \bibnamefont
  {Robinson}},\ }\href {\doibase 10.1088/0032-1028/11/10/010} {\bibfield
  {journal} {\bibinfo  {journal} {Plasma Physics}\ }\textbf {\bibinfo {volume}
  {11}},\ \bibinfo {pages} {893} (\bibinfo {year} {1969})}\BibitemShut
  {NoStop}%
\bibitem [{\citenamefont {{Bodin}}\ and\ \citenamefont
  {{Newton}}(1980)}]{Bodin1980}%
  \BibitemOpen
  \bibfield  {author} {\bibinfo {author} {\bibfnamefont {H.~A.~B.}\
  \bibnamefont {{Bodin}}}\ and\ \bibinfo {author} {\bibfnamefont {A.~A.}\
  \bibnamefont {{Newton}}},\ }\href@noop {} {\bibfield  {journal} {\bibinfo
  {journal} {Nuclear Fusion}\ }\textbf {\bibinfo {volume} {20}},\ \bibinfo
  {pages} {1255} (\bibinfo {year} {1980})}\BibitemShut {NoStop}%
\bibitem [{\citenamefont {Ortolani}\ and\ \citenamefont
  {Schnack}(1993)}]{Ortolani1993}%
  \BibitemOpen
  \bibfield  {author} {\bibinfo {author} {\bibfnamefont {S.}~\bibnamefont
  {Ortolani}}\ and\ \bibinfo {author} {\bibfnamefont {D.}~\bibnamefont
  {Schnack}},\ }\href {https://books.google.co.jp/books?id=WJoySA3lxe0C} {\emph
  {\bibinfo {title} {Magnetohydrodynamics of Plasma Relaxation}}}\ (\bibinfo
  {publisher} {World Scientific},\ \bibinfo {year} {1993})\BibitemShut
  {NoStop}%
\bibitem [{\citenamefont {Sarff}\ \emph {et~al.}(1997)\citenamefont {Sarff},
  \citenamefont {Lanier}, \citenamefont {Prager},\ and\ \citenamefont
  {Stoneking}}]{Sarff1997}%
  \BibitemOpen
  \bibfield  {author} {\bibinfo {author} {\bibfnamefont {J.~S.}\ \bibnamefont
  {Sarff}}, \bibinfo {author} {\bibfnamefont {N.~E.}\ \bibnamefont {Lanier}},
  \bibinfo {author} {\bibfnamefont {S.~C.}\ \bibnamefont {Prager}}, \ and\
  \bibinfo {author} {\bibfnamefont {M.~R.}\ \bibnamefont {Stoneking}},\ }\href
  {\doibase 10.1103/PhysRevLett.78.62} {\bibfield  {journal} {\bibinfo
  {journal} {Phys. Rev. Lett.}\ }\textbf {\bibinfo {volume} {78}},\ \bibinfo
  {pages} {62} (\bibinfo {year} {1997})}\BibitemShut {NoStop}%
\bibitem [{\citenamefont {Yagi}\ \emph {et~al.}(1999)\citenamefont {Yagi},
  \citenamefont {Sakakita}, \citenamefont {Shimada}, \citenamefont {Hayase},
  \citenamefont {Hirano}, \citenamefont {Hirota}, \citenamefont {Kiyama},
  \citenamefont {Koguchi}, \citenamefont {Maejima}, \citenamefont {Osakabe},
  \citenamefont {Sato}, \citenamefont {Sekine},\ and\ \citenamefont
  {Sugisaki}}]{Yagi1999}%
  \BibitemOpen
  \bibfield  {author} {\bibinfo {author} {\bibfnamefont {Y.}~\bibnamefont
  {Yagi}}, \bibinfo {author} {\bibfnamefont {H.}~\bibnamefont {Sakakita}},
  \bibinfo {author} {\bibfnamefont {T.}~\bibnamefont {Shimada}}, \bibinfo
  {author} {\bibfnamefont {K.}~\bibnamefont {Hayase}}, \bibinfo {author}
  {\bibfnamefont {Y.}~\bibnamefont {Hirano}}, \bibinfo {author} {\bibfnamefont
  {I.}~\bibnamefont {Hirota}}, \bibinfo {author} {\bibfnamefont
  {S.}~\bibnamefont {Kiyama}}, \bibinfo {author} {\bibfnamefont
  {H.}~\bibnamefont {Koguchi}}, \bibinfo {author} {\bibfnamefont
  {Y.}~\bibnamefont {Maejima}}, \bibinfo {author} {\bibfnamefont
  {T.}~\bibnamefont {Osakabe}}, \bibinfo {author} {\bibfnamefont
  {Y.}~\bibnamefont {Sato}}, \bibinfo {author} {\bibfnamefont {S.}~\bibnamefont
  {Sekine}}, \ and\ \bibinfo {author} {\bibfnamefont {K.}~\bibnamefont
  {Sugisaki}},\ }\href {\doibase 10.1088/0741-3335/41/2/009} {\bibfield
  {journal} {\bibinfo  {journal} {Plasma Physics and Controlled Fusion}\
  }\textbf {\bibinfo {volume} {41}},\ \bibinfo {pages} {255} (\bibinfo {year}
  {1999})}\BibitemShut {NoStop}%
\bibitem [{\citenamefont {Sarff}\ \emph {et~al.}(2003)\citenamefont {Sarff},
  \citenamefont {Almagri}, \citenamefont {Anderson}, \citenamefont {Biewer},
  \citenamefont {Blair}, \citenamefont {Cengher}, \citenamefont {Chapman},
  \citenamefont {Chattopadhyay}, \citenamefont {Craig}, \citenamefont
  {Den~Hartog} \emph {et~al.}}]{Sarff2003}%
  \BibitemOpen
  \bibfield  {author} {\bibinfo {author} {\bibfnamefont {J.}~\bibnamefont
  {Sarff}}, \bibinfo {author} {\bibfnamefont {A.}~\bibnamefont {Almagri}},
  \bibinfo {author} {\bibfnamefont {J.}~\bibnamefont {Anderson}}, \bibinfo
  {author} {\bibfnamefont {T.}~\bibnamefont {Biewer}}, \bibinfo {author}
  {\bibfnamefont {A.}~\bibnamefont {Blair}}, \bibinfo {author} {\bibfnamefont
  {M.}~\bibnamefont {Cengher}}, \bibinfo {author} {\bibfnamefont
  {B.}~\bibnamefont {Chapman}}, \bibinfo {author} {\bibfnamefont
  {P.}~\bibnamefont {Chattopadhyay}}, \bibinfo {author} {\bibfnamefont
  {D.}~\bibnamefont {Craig}}, \bibinfo {author} {\bibfnamefont
  {D.}~\bibnamefont {Den~Hartog}},  \emph {et~al.},\ }\href@noop {} {\bibfield
  {journal} {\bibinfo  {journal} {Nuclear Fusion}\ }\textbf {\bibinfo {volume}
  {43}},\ \bibinfo {pages} {1684} (\bibinfo {year} {2003})}\BibitemShut
  {NoStop}%
\bibitem [{\citenamefont {Ding}\ \emph {et~al.}(2004)\citenamefont {Ding},
  \citenamefont {Brower}, \citenamefont {Craig}, \citenamefont {Deng},
  \citenamefont {Fiksel}, \citenamefont {Mirnov}, \citenamefont {Prager},
  \citenamefont {Sarff},\ and\ \citenamefont {Svidzinski}}]{Ding2004}%
  \BibitemOpen
  \bibfield  {author} {\bibinfo {author} {\bibfnamefont {W.~X.}\ \bibnamefont
  {Ding}}, \bibinfo {author} {\bibfnamefont {D.~L.}\ \bibnamefont {Brower}},
  \bibinfo {author} {\bibfnamefont {D.}~\bibnamefont {Craig}}, \bibinfo
  {author} {\bibfnamefont {B.~H.}\ \bibnamefont {Deng}}, \bibinfo {author}
  {\bibfnamefont {G.}~\bibnamefont {Fiksel}}, \bibinfo {author} {\bibfnamefont
  {V.}~\bibnamefont {Mirnov}}, \bibinfo {author} {\bibfnamefont {S.~C.}\
  \bibnamefont {Prager}}, \bibinfo {author} {\bibfnamefont {J.~S.}\
  \bibnamefont {Sarff}}, \ and\ \bibinfo {author} {\bibfnamefont
  {V.}~\bibnamefont {Svidzinski}},\ }\href {\doibase
  10.1103/PhysRevLett.93.045002} {\bibfield  {journal} {\bibinfo  {journal}
  {Phys. Rev. Lett.}\ }\textbf {\bibinfo {volume} {93}},\ \bibinfo {pages}
  {045002} (\bibinfo {year} {2004})}\BibitemShut {NoStop}%
\bibitem [{\citenamefont {Prager}\ \emph {et~al.}(2005)\citenamefont {Prager},
  \citenamefont {Adney}, \citenamefont {Almagri}, \citenamefont {Anderson},
  \citenamefont {Blair}, \citenamefont {Brower}, \citenamefont {Cengher},
  \citenamefont {Chapman}, \citenamefont {Choi}, \citenamefont {Craig} \emph
  {et~al.}}]{Prager2005}%
  \BibitemOpen
  \bibfield  {author} {\bibinfo {author} {\bibfnamefont {S.}~\bibnamefont
  {Prager}}, \bibinfo {author} {\bibfnamefont {J.}~\bibnamefont {Adney}},
  \bibinfo {author} {\bibfnamefont {A.}~\bibnamefont {Almagri}}, \bibinfo
  {author} {\bibfnamefont {J.}~\bibnamefont {Anderson}}, \bibinfo {author}
  {\bibfnamefont {A.}~\bibnamefont {Blair}}, \bibinfo {author} {\bibfnamefont
  {D.}~\bibnamefont {Brower}}, \bibinfo {author} {\bibfnamefont
  {M.}~\bibnamefont {Cengher}}, \bibinfo {author} {\bibfnamefont
  {B.}~\bibnamefont {Chapman}}, \bibinfo {author} {\bibfnamefont
  {S.}~\bibnamefont {Choi}}, \bibinfo {author} {\bibfnamefont {D.}~\bibnamefont
  {Craig}},  \emph {et~al.},\ }\href@noop {} {\bibfield  {journal} {\bibinfo
  {journal} {Nuclear Fusion}\ }\textbf {\bibinfo {volume} {45}},\ \bibinfo
  {pages} {S276} (\bibinfo {year} {2005})}\BibitemShut {NoStop}%
\bibitem [{\citenamefont {Lorenzini}\ \emph {et~al.}(2009)\citenamefont
  {Lorenzini}, \citenamefont {Martines}, \citenamefont {Piovesan},
  \citenamefont {Terranova}, \citenamefont {Zanca}, \citenamefont {Zuin},
  \citenamefont {Alfier}, \citenamefont {Bonfiglio}, \citenamefont {Bonomo},
  \citenamefont {Canton} \emph {et~al.}}]{Lorenzini2009}%
  \BibitemOpen
  \bibfield  {author} {\bibinfo {author} {\bibfnamefont {R.}~\bibnamefont
  {Lorenzini}}, \bibinfo {author} {\bibfnamefont {E.}~\bibnamefont {Martines}},
  \bibinfo {author} {\bibfnamefont {P.}~\bibnamefont {Piovesan}}, \bibinfo
  {author} {\bibfnamefont {D.}~\bibnamefont {Terranova}}, \bibinfo {author}
  {\bibfnamefont {P.}~\bibnamefont {Zanca}}, \bibinfo {author} {\bibfnamefont
  {M.}~\bibnamefont {Zuin}}, \bibinfo {author} {\bibfnamefont {A.}~\bibnamefont
  {Alfier}}, \bibinfo {author} {\bibfnamefont {D.}~\bibnamefont {Bonfiglio}},
  \bibinfo {author} {\bibfnamefont {F.}~\bibnamefont {Bonomo}}, \bibinfo
  {author} {\bibfnamefont {A.}~\bibnamefont {Canton}},  \emph {et~al.},\
  }\href@noop {} {\bibfield  {journal} {\bibinfo  {journal} {Nature Physics}\
  }\textbf {\bibinfo {volume} {5}},\ \bibinfo {pages} {570} (\bibinfo {year}
  {2009})}\BibitemShut {NoStop}%
\bibitem [{\citenamefont {{Beltrami}}(1889)}]{Beltrami1889}%
  \BibitemOpen
  \bibfield  {author} {\bibinfo {author} {\bibfnamefont {E.}~\bibnamefont
  {{Beltrami}}},\ }\href {\doibase 10.1007/BF02719090} {\bibfield  {journal}
  {\bibinfo  {journal} {Il Nuovo Cimento}\ }\textbf {\bibinfo {volume} {25}},\
  \bibinfo {pages} {212} (\bibinfo {year} {1889})}\BibitemShut {NoStop}%
\bibitem [{\citenamefont {{L{\"u}st}}\ and\ \citenamefont
  {{Schl{\"u}ter}}(1954)}]{Lust1954}%
  \BibitemOpen
  \bibfield  {author} {\bibinfo {author} {\bibfnamefont {R.}~\bibnamefont
  {{L{\"u}st}}}\ and\ \bibinfo {author} {\bibfnamefont {A.}~\bibnamefont
  {{Schl{\"u}ter}}},\ }\href@noop {} {\bibfield  {journal} {\bibinfo  {journal}
  {Zeitschrift f{\"u}r Astrophysik}\ }\textbf {\bibinfo {volume} {34}},\
  \bibinfo {pages} {263} (\bibinfo {year} {1954})}\BibitemShut {NoStop}%
\bibitem [{\citenamefont {{Chandrasekhar}}\ and\ \citenamefont
  {{Kendall}}(1957)}]{Chandrasekhar1957}%
  \BibitemOpen
  \bibfield  {author} {\bibinfo {author} {\bibfnamefont {S.}~\bibnamefont
  {{Chandrasekhar}}}\ and\ \bibinfo {author} {\bibfnamefont {P.~C.}\
  \bibnamefont {{Kendall}}},\ }\href {\doibase 10.1086/146413} {\bibfield
  {journal} {\bibinfo  {journal} {\apj}\ }\textbf {\bibinfo {volume} {126}},\
  \bibinfo {pages} {457} (\bibinfo {year} {1957})}\BibitemShut {NoStop}%
\bibitem [{\citenamefont {{Chandrasekhar}}\ and\ \citenamefont
  {{Woltjer}}(1958)}]{Chandrasekhar1958}%
  \BibitemOpen
  \bibfield  {author} {\bibinfo {author} {\bibfnamefont {S.}~\bibnamefont
  {{Chandrasekhar}}}\ and\ \bibinfo {author} {\bibfnamefont {L.}~\bibnamefont
  {{Woltjer}}},\ }\href {\doibase 10.1073/pnas.44.4.285} {\bibfield  {journal}
  {\bibinfo  {journal} {Proceedings of the National Academy of Science}\
  }\textbf {\bibinfo {volume} {44}},\ \bibinfo {pages} {285} (\bibinfo {year}
  {1958})}\BibitemShut {NoStop}%
\bibitem [{\citenamefont {{Woltjer}}(1958)}]{Woltjer1958}%
  \BibitemOpen
  \bibfield  {author} {\bibinfo {author} {\bibfnamefont {L.}~\bibnamefont
  {{Woltjer}}},\ }\href {\doibase 10.1073/pnas.44.6.489} {\bibfield  {journal}
  {\bibinfo  {journal} {Proceedings of the National Academy of Science}\
  }\textbf {\bibinfo {volume} {44}},\ \bibinfo {pages} {489} (\bibinfo {year}
  {1958})}\BibitemShut {NoStop}%
\bibitem [{\citenamefont {Wells}\ and\ \citenamefont
  {Norwood}(1969)}]{Wells1969}%
  \BibitemOpen
  \bibfield  {author} {\bibinfo {author} {\bibfnamefont {D.~R.}\ \bibnamefont
  {Wells}}\ and\ \bibinfo {author} {\bibfnamefont {J.}~\bibnamefont
  {Norwood}},\ }\href {\doibase 10.1017/S0022377800004165} {\bibfield
  {journal} {\bibinfo  {journal} {Journal of Plasma Physics}\ }\textbf
  {\bibinfo {volume} {3}},\ \bibinfo {pages} {21–46} (\bibinfo {year}
  {1969})}\BibitemShut {NoStop}%
\bibitem [{\citenamefont {{Moffatt}}(1978)}]{Moffatt1978}%
  \BibitemOpen
  \bibfield  {author} {\bibinfo {author} {\bibfnamefont {H.~K.}\ \bibnamefont
  {{Moffatt}}},\ }\href {https://books.google.co.jp/books?id=cAo4AAAAIAAJ}
  {\emph {\bibinfo {title} {Magnetic Field Generation in Electrically
  Conducting Fluids}}},\ Cambridge Monographs on Mechanics\ (\bibinfo
  {publisher} {Cambridge University Press},\ \bibinfo {year}
  {1978})\BibitemShut {NoStop}%
\bibitem [{\citenamefont {Berger}\ and\ \citenamefont
  {Field}(1984)}]{Berger1984}%
  \BibitemOpen
  \bibfield  {author} {\bibinfo {author} {\bibfnamefont {M.}~\bibnamefont
  {Berger}}\ and\ \bibinfo {author} {\bibfnamefont {G.}~\bibnamefont {Field}},\
  }\href {\doibase 10.1017/S0022112084002019} {\bibfield  {journal} {\bibinfo
  {journal} {Journal of Fluid Mechanics}\ }\textbf {\bibinfo {volume} {147}},\
  \bibinfo {pages} {133–148} (\bibinfo {year} {1984})}\BibitemShut {NoStop}%
\bibitem [{\citenamefont {Arnold}\ and\ \citenamefont
  {Khesin}(1999)}]{Arnold1999}%
  \BibitemOpen
  \bibfield  {author} {\bibinfo {author} {\bibfnamefont {V.}~\bibnamefont
  {Arnold}}\ and\ \bibinfo {author} {\bibfnamefont {B.}~\bibnamefont
  {Khesin}},\ }\href {https://books.google.co.jp/books?id=9Iwrt0l0nFMC} {\emph
  {\bibinfo {title} {Topological Methods in Hydrodynamics}}},\ Applied
  Mathematical Sciences\ (\bibinfo  {publisher} {Springer, New York},\ \bibinfo
  {year} {1999})\BibitemShut {NoStop}%
\bibitem [{\citenamefont {Taylor}(1974)}]{Taylor1974}%
  \BibitemOpen
  \bibfield  {author} {\bibinfo {author} {\bibfnamefont {J.~B.}\ \bibnamefont
  {Taylor}},\ }\href {\doibase 10.1103/PhysRevLett.33.1139} {\bibfield
  {journal} {\bibinfo  {journal} {Phys. Rev. Lett.}\ }\textbf {\bibinfo
  {volume} {33}},\ \bibinfo {pages} {1139} (\bibinfo {year}
  {1974})}\BibitemShut {NoStop}%
\bibitem [{\citenamefont {Taylor}(1986)}]{Taylor1986}%
  \BibitemOpen
  \bibfield  {author} {\bibinfo {author} {\bibfnamefont {J.~B.}\ \bibnamefont
  {Taylor}},\ }\href {\doibase 10.1103/RevModPhys.58.741} {\bibfield  {journal}
  {\bibinfo  {journal} {Rev. Mod. Phys.}\ }\textbf {\bibinfo {volume} {58}},\
  \bibinfo {pages} {741} (\bibinfo {year} {1986})}\BibitemShut {NoStop}%
\bibitem [{\citenamefont {Schnack}(2009)}]{Schnack2009}%
  \BibitemOpen
  \bibfield  {author} {\bibinfo {author} {\bibfnamefont {D.}~\bibnamefont
  {Schnack}},\ }\href {https://books.google.co.jp/books?id=Ebon7NTbL0EC} {\emph
  {\bibinfo {title} {Lectures in Magnetohydrodynamics: With an Appendix on
  Extended MHD}}},\ Lecture Notes in Physics\ (\bibinfo  {publisher} {Springer,
  Berlin, Heidelberg},\ \bibinfo {year} {2009})\BibitemShut {NoStop}%
\bibitem [{\citenamefont {Caramana}\ \emph {et~al.}(1983)\citenamefont
  {Caramana}, \citenamefont {Nebel},\ and\ \citenamefont
  {Schnack}}]{Caramana1983}%
  \BibitemOpen
  \bibfield  {author} {\bibinfo {author} {\bibfnamefont {E.}~\bibnamefont
  {Caramana}}, \bibinfo {author} {\bibfnamefont {R.}~\bibnamefont {Nebel}}, \
  and\ \bibinfo {author} {\bibfnamefont {D.}~\bibnamefont {Schnack}},\
  }\href@noop {} {\bibfield  {journal} {\bibinfo  {journal} {The Physics of
  Fluids}\ }\textbf {\bibinfo {volume} {26}},\ \bibinfo {pages} {1305}
  (\bibinfo {year} {1983})}\BibitemShut {NoStop}%
\bibitem [{\citenamefont {Schnack}\ \emph {et~al.}(1985)\citenamefont
  {Schnack}, \citenamefont {Caramana},\ and\ \citenamefont
  {Nebel}}]{Schnack1985}%
  \BibitemOpen
  \bibfield  {author} {\bibinfo {author} {\bibfnamefont {D.}~\bibnamefont
  {Schnack}}, \bibinfo {author} {\bibfnamefont {E.}~\bibnamefont {Caramana}}, \
  and\ \bibinfo {author} {\bibfnamefont {R.}~\bibnamefont {Nebel}},\
  }\href@noop {} {\bibfield  {journal} {\bibinfo  {journal} {The Physics of
  Fluids}\ }\textbf {\bibinfo {volume} {28}},\ \bibinfo {pages} {321} (\bibinfo
  {year} {1985})}\BibitemShut {NoStop}%
\bibitem [{\citenamefont {Strauss}(1985)}]{Strauss1985}%
  \BibitemOpen
  \bibfield  {author} {\bibinfo {author} {\bibfnamefont {H.}~\bibnamefont
  {Strauss}},\ }\href@noop {} {\bibfield  {journal} {\bibinfo  {journal} {The
  Physics of Fluids}\ }\textbf {\bibinfo {volume} {28}},\ \bibinfo {pages}
  {2786} (\bibinfo {year} {1985})}\BibitemShut {NoStop}%
\bibitem [{\citenamefont {Kusano}\ and\ \citenamefont
  {Sato}(1987)}]{Kusano1987}%
  \BibitemOpen
  \bibfield  {author} {\bibinfo {author} {\bibfnamefont {K.}~\bibnamefont
  {Kusano}}\ and\ \bibinfo {author} {\bibfnamefont {T.}~\bibnamefont {Sato}},\
  }\href@noop {} {\bibfield  {journal} {\bibinfo  {journal} {Nuclear Fusion}\
  }\textbf {\bibinfo {volume} {27}},\ \bibinfo {pages} {821} (\bibinfo {year}
  {1987})}\BibitemShut {NoStop}%
\bibitem [{\citenamefont {Holmes}\ \emph {et~al.}(1988)\citenamefont {Holmes},
  \citenamefont {Carreras}, \citenamefont {Diamond},\ and\ \citenamefont
  {Lynch}}]{Holmes1988}%
  \BibitemOpen
  \bibfield  {author} {\bibinfo {author} {\bibfnamefont {J.}~\bibnamefont
  {Holmes}}, \bibinfo {author} {\bibfnamefont {B.}~\bibnamefont {Carreras}},
  \bibinfo {author} {\bibfnamefont {P.}~\bibnamefont {Diamond}}, \ and\
  \bibinfo {author} {\bibfnamefont {V.~E.}\ \bibnamefont {Lynch}},\ }\href@noop
  {} {\bibfield  {journal} {\bibinfo  {journal} {The Physics of Fluids}\
  }\textbf {\bibinfo {volume} {31}},\ \bibinfo {pages} {1166} (\bibinfo {year}
  {1988})}\BibitemShut {NoStop}%
\bibitem [{\citenamefont {Ho}\ and\ \citenamefont {Craddock}(1991)}]{Ho1991}%
  \BibitemOpen
  \bibfield  {author} {\bibinfo {author} {\bibfnamefont {Y.}~\bibnamefont
  {Ho}}\ and\ \bibinfo {author} {\bibfnamefont {G.}~\bibnamefont {Craddock}},\
  }\href@noop {} {\bibfield  {journal} {\bibinfo  {journal} {Physics of Fluids
  B: Plasma Physics}\ }\textbf {\bibinfo {volume} {3}},\ \bibinfo {pages} {721}
  (\bibinfo {year} {1991})}\BibitemShut {NoStop}%
\bibitem [{\citenamefont {Diamond}\ and\ \citenamefont
  {Malkov}(2003)}]{Diamond2003}%
  \BibitemOpen
  \bibfield  {author} {\bibinfo {author} {\bibfnamefont {P.}~\bibnamefont
  {Diamond}}\ and\ \bibinfo {author} {\bibfnamefont {M.}~\bibnamefont
  {Malkov}},\ }\href@noop {} {\bibfield  {journal} {\bibinfo  {journal}
  {Physics of Plasmas}\ }\textbf {\bibinfo {volume} {10}},\ \bibinfo {pages}
  {2322} (\bibinfo {year} {2003})}\BibitemShut {NoStop}%
\bibitem [{\citenamefont {Marrelli}\ \emph {et~al.}(2005)\citenamefont
  {Marrelli}, \citenamefont {Frassinetti}, \citenamefont {Martin},
  \citenamefont {Craig},\ and\ \citenamefont {Sarff}}]{Marrelli2005}%
  \BibitemOpen
  \bibfield  {author} {\bibinfo {author} {\bibfnamefont {L.}~\bibnamefont
  {Marrelli}}, \bibinfo {author} {\bibfnamefont {L.}~\bibnamefont
  {Frassinetti}}, \bibinfo {author} {\bibfnamefont {P.}~\bibnamefont {Martin}},
  \bibinfo {author} {\bibfnamefont {D.}~\bibnamefont {Craig}}, \ and\ \bibinfo
  {author} {\bibfnamefont {J.}~\bibnamefont {Sarff}},\ }\href@noop {}
  {\bibfield  {journal} {\bibinfo  {journal} {Physics of Plasmas}\ }\textbf
  {\bibinfo {volume} {12}},\ \bibinfo {pages} {030701} (\bibinfo {year}
  {2005})}\BibitemShut {NoStop}%
\bibitem [{\citenamefont {Bhattacharjee}\ \emph {et~al.}(1980)\citenamefont
  {Bhattacharjee}, \citenamefont {Dewar},\ and\ \citenamefont
  {Monticello}}]{Bhattacharjee1980}%
  \BibitemOpen
  \bibfield  {author} {\bibinfo {author} {\bibfnamefont {A.}~\bibnamefont
  {Bhattacharjee}}, \bibinfo {author} {\bibfnamefont {R.~L.}\ \bibnamefont
  {Dewar}}, \ and\ \bibinfo {author} {\bibfnamefont {D.~A.}\ \bibnamefont
  {Monticello}},\ }\href {\doibase 10.1103/PhysRevLett.45.347} {\bibfield
  {journal} {\bibinfo  {journal} {Phys. Rev. Lett.}\ }\textbf {\bibinfo
  {volume} {45}},\ \bibinfo {pages} {347} (\bibinfo {year} {1980})}\BibitemShut
  {NoStop}%
\bibitem [{\citenamefont {Bhattacharjee}\ and\ \citenamefont
  {Dewar}(1982)}]{Bhattacharjee1982}%
  \BibitemOpen
  \bibfield  {author} {\bibinfo {author} {\bibfnamefont {A.}~\bibnamefont
  {Bhattacharjee}}\ and\ \bibinfo {author} {\bibfnamefont {R.~L.}\ \bibnamefont
  {Dewar}},\ }\href@noop {} {\bibfield  {journal} {\bibinfo  {journal} {The
  Physics of Fluids}\ }\textbf {\bibinfo {volume} {25}},\ \bibinfo {pages}
  {887} (\bibinfo {year} {1982})}\BibitemShut {NoStop}%
\bibitem [{\citenamefont {Qin}\ \emph {et~al.}(2012)\citenamefont {Qin},
  \citenamefont {Liu}, \citenamefont {Li},\ and\ \citenamefont
  {Squire}}]{Qin2012}%
  \BibitemOpen
  \bibfield  {author} {\bibinfo {author} {\bibfnamefont {H.}~\bibnamefont
  {Qin}}, \bibinfo {author} {\bibfnamefont {W.}~\bibnamefont {Liu}}, \bibinfo
  {author} {\bibfnamefont {H.}~\bibnamefont {Li}}, \ and\ \bibinfo {author}
  {\bibfnamefont {J.}~\bibnamefont {Squire}},\ }\href {\doibase
  10.1103/PhysRevLett.109.235001} {\bibfield  {journal} {\bibinfo  {journal}
  {Phys. Rev. Lett.}\ }\textbf {\bibinfo {volume} {109}},\ \bibinfo {pages}
  {235001} (\bibinfo {year} {2012})}\BibitemShut {NoStop}%
\bibitem [{\citenamefont {Hirono}\ \emph {et~al.}(2015)\citenamefont {Hirono},
  \citenamefont {Kharzeev},\ and\ \citenamefont {Yin}}]{Hirono:2015rla}%
  \BibitemOpen
  \bibfield  {author} {\bibinfo {author} {\bibfnamefont {Y.}~\bibnamefont
  {Hirono}}, \bibinfo {author} {\bibfnamefont {D.}~\bibnamefont {Kharzeev}}, \
  and\ \bibinfo {author} {\bibfnamefont {Y.}~\bibnamefont {Yin}},\ }\href
  {\doibase 10.1103/PhysRevD.92.125031} {\bibfield  {journal} {\bibinfo
  {journal} {Phys. Rev. D}\ }\textbf {\bibinfo {volume} {92}},\ \bibinfo
  {pages} {125031} (\bibinfo {year} {2015})},\ \Eprint
  {http://arxiv.org/abs/1509.07790} {arXiv:1509.07790 [hep-th]} \BibitemShut
  {NoStop}%
\bibitem [{\citenamefont {Vilenkin}(1980)}]{Vilenkin:1980fu}%
  \BibitemOpen
  \bibfield  {author} {\bibinfo {author} {\bibfnamefont {A.}~\bibnamefont
  {Vilenkin}},\ }\href {\doibase 10.1103/PhysRevD.22.3080} {\bibfield
  {journal} {\bibinfo  {journal} {Phys. Rev. D}\ }\textbf {\bibinfo {volume}
  {22}},\ \bibinfo {pages} {3080} (\bibinfo {year} {1980})}\BibitemShut
  {NoStop}%
\bibitem [{\citenamefont {Kharzeev}\ \emph {et~al.}(2008)\citenamefont
  {Kharzeev}, \citenamefont {McLerran},\ and\ \citenamefont
  {Warringa}}]{Kharzeev:2007jp}%
  \BibitemOpen
  \bibfield  {author} {\bibinfo {author} {\bibfnamefont {D.~E.}\ \bibnamefont
  {Kharzeev}}, \bibinfo {author} {\bibfnamefont {L.~D.}\ \bibnamefont
  {McLerran}}, \ and\ \bibinfo {author} {\bibfnamefont {H.~J.}\ \bibnamefont
  {Warringa}},\ }\href {\doibase 10.1016/j.nuclphysa.2008.02.298} {\bibfield
  {journal} {\bibinfo  {journal} {Nucl. Phys. A}\ }\textbf {\bibinfo {volume}
  {803}},\ \bibinfo {pages} {227} (\bibinfo {year} {2008})},\ \Eprint
  {http://arxiv.org/abs/0711.0950} {arXiv:0711.0950 [hep-ph]} \BibitemShut
  {NoStop}%
\bibitem [{\citenamefont {Fukushima}\ \emph {et~al.}(2008)\citenamefont
  {Fukushima}, \citenamefont {Kharzeev},\ and\ \citenamefont
  {Warringa}}]{Fukushima:2008xe}%
  \BibitemOpen
  \bibfield  {author} {\bibinfo {author} {\bibfnamefont {K.}~\bibnamefont
  {Fukushima}}, \bibinfo {author} {\bibfnamefont {D.~E.}\ \bibnamefont
  {Kharzeev}}, \ and\ \bibinfo {author} {\bibfnamefont {H.~J.}\ \bibnamefont
  {Warringa}},\ }\href {\doibase 10.1103/PhysRevD.78.074033} {\bibfield
  {journal} {\bibinfo  {journal} {Phys. Rev. D}\ }\textbf {\bibinfo {volume}
  {78}},\ \bibinfo {pages} {074033} (\bibinfo {year} {2008})},\ \Eprint
  {http://arxiv.org/abs/0808.3382} {arXiv:0808.3382 [hep-ph]} \BibitemShut
  {NoStop}%
\bibitem [{\citenamefont {Kharzeev}\ \emph {et~al.}(2013)\citenamefont
  {Kharzeev}, \citenamefont {Landsteiner}, \citenamefont {Schmitt},\ and\
  \citenamefont {Yee}}]{Kharzeev:2012ph}%
  \BibitemOpen
  \bibfield  {author} {\bibinfo {author} {\bibfnamefont {D.~E.}\ \bibnamefont
  {Kharzeev}}, \bibinfo {author} {\bibfnamefont {K.}~\bibnamefont
  {Landsteiner}}, \bibinfo {author} {\bibfnamefont {A.}~\bibnamefont
  {Schmitt}}, \ and\ \bibinfo {author} {\bibfnamefont {H.-U.}\ \bibnamefont
  {Yee}},\ }\href {\doibase 10.1007/978-3-642-37305-3_1} {\bibfield  {journal}
  {\bibinfo  {journal} {Lect. Notes Phys.}\ }\textbf {\bibinfo {volume}
  {871}},\ \bibinfo {pages} {1} (\bibinfo {year} {2013})},\ \Eprint
  {http://arxiv.org/abs/1211.6245} {arXiv:1211.6245 [hep-ph]} \BibitemShut
  {NoStop}%
\bibitem [{\citenamefont {Kharzeev}\ \emph {et~al.}(2016)\citenamefont
  {Kharzeev}, \citenamefont {Liao}, \citenamefont {Voloshin},\ and\
  \citenamefont {Wang}}]{Kharzeev:2015znc}%
  \BibitemOpen
  \bibfield  {author} {\bibinfo {author} {\bibfnamefont {D.~E.}\ \bibnamefont
  {Kharzeev}}, \bibinfo {author} {\bibfnamefont {J.}~\bibnamefont {Liao}},
  \bibinfo {author} {\bibfnamefont {S.~A.}\ \bibnamefont {Voloshin}}, \ and\
  \bibinfo {author} {\bibfnamefont {G.}~\bibnamefont {Wang}},\ }\href {\doibase
  10.1016/j.ppnp.2016.01.001} {\bibfield  {journal} {\bibinfo  {journal} {Prog.
  Part. Nucl. Phys.}\ }\textbf {\bibinfo {volume} {88}},\ \bibinfo {pages} {1}
  (\bibinfo {year} {2016})},\ \Eprint {http://arxiv.org/abs/1511.04050}
  {arXiv:1511.04050 [hep-ph]} \BibitemShut {NoStop}%
\bibitem [{\citenamefont {Xia}\ \emph {et~al.}(2016)\citenamefont {Xia},
  \citenamefont {Qin},\ and\ \citenamefont {Wang}}]{Xia:2016any}%
  \BibitemOpen
  \bibfield  {author} {\bibinfo {author} {\bibfnamefont {X.-l.}\ \bibnamefont
  {Xia}}, \bibinfo {author} {\bibfnamefont {H.}~\bibnamefont {Qin}}, \ and\
  \bibinfo {author} {\bibfnamefont {Q.}~\bibnamefont {Wang}},\ }\href {\doibase
  10.1103/PhysRevD.94.054042} {\bibfield  {journal} {\bibinfo  {journal} {Phys.
  Rev. D}\ }\textbf {\bibinfo {volume} {94}},\ \bibinfo {pages} {054042}
  (\bibinfo {year} {2016})},\ \Eprint {http://arxiv.org/abs/1607.01126}
  {arXiv:1607.01126 [nucl-th]} \BibitemShut {NoStop}%
\bibitem [{\citenamefont {Chen}\ and\ \citenamefont {Fan}(2013)}]{Chen2013}%
  \BibitemOpen
  \bibfield  {author} {\bibinfo {author} {\bibfnamefont {J.-h.}\ \bibnamefont
  {Chen}}\ and\ \bibinfo {author} {\bibfnamefont {H.-y.}\ \bibnamefont {Fan}},\
  }\href {\doibase 10.1103/PhysRevLett.110.269501} {\bibfield  {journal}
  {\bibinfo  {journal} {Phys. Rev. Lett.}\ }\textbf {\bibinfo {volume} {110}},\
  \bibinfo {pages} {269501} (\bibinfo {year} {2013})}\BibitemShut {NoStop}%
\bibitem [{\citenamefont {Poincar{\'e}}(1890)}]{Poincare1890}%
  \BibitemOpen
  \bibfield  {author} {\bibinfo {author} {\bibfnamefont {H.}~\bibnamefont
  {Poincar{\'e}}},\ }\href@noop {} {\bibfield  {journal} {\bibinfo  {journal}
  {American Journal of Mathematics}\ }\textbf {\bibinfo {volume} {12}},\
  \bibinfo {pages} {211} (\bibinfo {year} {1890})}\BibitemShut {NoStop}%
\bibitem [{\citenamefont {Jackson}(1999)}]{Jackson1999}%
  \BibitemOpen
  \bibfield  {author} {\bibinfo {author} {\bibfnamefont {J.~D.}\ \bibnamefont
  {Jackson}},\ }\href {http://cdsweb.cern.ch/record/490457} {\emph {\bibinfo
  {title} {Classical electrodynamics}}},\ \bibinfo {edition} {3rd}\ ed.\
  (\bibinfo  {publisher} {Wiley},\ \bibinfo {address} {New York},\ \bibinfo
  {year} {1999})\BibitemShut {NoStop}%
\end{thebibliography}%

\onecolumngrid

\appendix

\section{Proof of Eq. (\ref{eq:wq-h2-continuous})}

\label{sec:proof}Using the asymptotic series of the incomplete gamma
function $\Gamma(s,z)$ when $z\rightarrow\infty$ we obtain 
\begin{eqnarray}
\int_{a}^{\infty}k^{p}e^{-2k^{2}\eta t}dk & = & \frac{1}{2}(2\eta t)^{-(p+1)/2}\Gamma\left(\frac{1+p}{2},2a^{2}\eta t\right)\nonumber \\
 & \sim & e^{-2a^{2}\eta t}\sum_{j=1}^{\infty}\frac{a^{p-(2j-1)}}{(4\eta t)^{j}}\prod_{h=1}^{j-1}\left[p-\left(2h-1\right)\right],\label{eq:expansion}
\end{eqnarray}
where the index $j$ is called the integral approximation order. The
integration of the $i$-th order term of Taylor expansion of $f(k)$
is evaluated as 
\begin{align}
X_{i} & =\int_{a}^{\infty}k^{n}\left[\frac{1}{i!}(k-a)^{i}f^{(i)}(a)\right]e^{-2k^{2}\eta t}dk\nonumber \\
 & =\frac{1}{i!}f^{(i)}(a)\sum_{m=0}^{i}C_{i}^{m}\left(\int_{a}^{\infty}k^{n+m}e^{-2k^{2}\eta t}dk\right)a^{i-m}(-1)^{i-m},
\end{align}
where $n=0,1,2$ for $Q,H,W$ respectively. 

Now we consider the $j$-th order term in $X_{i}$ with the expansion
of $\int_{a}^{\infty}k^{n+m}e^{-2k^{2}\eta t}dk$ following Eq. (\ref{eq:expansion}),
\begin{eqnarray}
X_{ij} & = & \frac{1}{i!}f^{(i)}(a)\sum_{m=0}^{i}C_{i}^{m}e^{-2a^{2}\eta t}\frac{a^{n+m-(2j-1)}}{(4\eta t)^{j}}\nonumber \\
 &  & \times\prod_{h=1}^{j-1}\left[n+m-(2h-1)\right]a^{i-m}(-1)^{i-m}\nonumber \\
 & = & \frac{1}{i!}f^{(i)}(a)e^{-2a^{2}\eta t}\frac{a^{n+i-(2j-1)}}{(4\eta t)^{j}}\nonumber \\
 &  & \times\sum_{m=0}^{i}C_{i}^{m}(-1)^{i-m}\left(\prod_{h=1}^{j-1}\left[n+m-(2h-1)\right]\right)\nonumber \\
 & = & \frac{1}{i!}f^{(i)}(a)e^{-2a^{2}\eta t}\frac{a^{n+i-(2j-1)}}{(4\eta t)^{j}}\sum_{m=0}^{j-1}i!S(m,i)q_{m}(j,n).
\end{eqnarray}
Here we have used the second kind Stirling number $S(m,i)$, with
the function $q$ being defined as 
\begin{equation}
\prod_{h=1}^{j-1}\left[n+m-(2h-1)\right]=\sum_{i^{\prime}=0}^{j-1}q_{i^{\prime}}(j,n)m^{i^{\prime}}.
\end{equation}
Since $S(m,i)=0$ for $m<i$, the lowest order term of $1/t$ among
$X_{ij}$ must require $j=i+1$ with the lowest $i$-th non-zero derivative
($i\geq0$), which is 
\begin{equation}
f^{(i)}(a)e^{-2a^{2}\eta t}\frac{a^{n-i-1}}{(4\eta t)^{i+1}}.
\end{equation}
And then as $t$ goes to infinity, the leading term of $WQ/H^{2}$
becomes 
\begin{align}
\frac{WQ}{H^{2}} & \sim\frac{\phi^{(i)}(a)e^{-2a^{2}\eta t}\frac{a^{1-i}}{(4\eta t)^{i+1}}\phi^{(i)}(a)e^{-2a^{2}\eta t}\frac{a^{-1-i}}{(4\eta t)^{i+1}}}{\varphi^{(i)}(a)e^{-2a^{2}\eta t}\frac{a^{-i}}{(4\eta t)^{i+1}}\varphi^{(i)}(a)e^{-2a^{2}\eta t}\frac{a^{-i}}{(4\eta t)^{i+1}}}\nonumber \\
 & =\left[\frac{\phi^{(i)}(a)}{\varphi^{(i)}(a)}\right]^{2},
\end{align}
where $\phi^{(i)}(a)$ and $\varphi^{(i)}(a)$ are the same $i$-th
order derivatives of $\phi(k)$ and $\varphi(k)$ defined in Eq. (\ref{eq:phi-beta})
at $k=a$.
\end{document}